\documentclass{aa}  
\usepackage{graphicx}
\usepackage{newtxtext,newtxmath}
\usepackage{xcolor}
\usepackage[]{changes}

\definechangesauthor[name=Dominique,color=orange]{dom}
\definechangesauthor[name=Julien]{jul}
\newcommand{\cmfast}{{\texttt{21cmFAST} }}

\newcommand{\zo}{z_{\mathrm{obs}}}
\newcommand{\ZR}{z_{\mathrm{reion}}(\Vec{r})}
\newcommand{\TR}{t_{\mathrm{reion}}(\Vec{r})}
\newcommand{\mpc}{\mathrm{cMpc/h}}
\newcommand{\hmpc}{\mathrm{h/cMpc}}
\usepackage{diagbox}
\usepackage{enumitem}

\usepackage{booktabs}
\usepackage{siunitx}
\usepackage{multirow}
\usepackage{marvosym}
\usepackage[utf8]{inputenc}

\usepackage{tikz}
\usetikzlibrary{arrows.meta, positioning}

\usepackage{hyperref} 

\usepackage{titlesec}
\titlespacing*{\subsection}{0pt}{*1}{*0.5}

\AtBeginDocument{\hypersetup{pdfborder={0 0 1}}}
\usepackage{pdftexcmds}
\DeclareFontFamily{U}{cbgreek}{}
\DeclareFontShape{U}{cbgreek}{m}{n}{
        <-6>    grmn0500
        <6-7>   grmn0600
        <7-8>   grmn0700
        <8-9>   grmn0800
        <9-10>  grmn0900
        <10-12> grmn1000
        <12-17> grmn1200
        <17->   grmn1728
      }{}
\DeclareFontShape{U}{cbgreek}{bx}{n}{
        <-6>    grxn0500
        <6-7>   grxn0600
        <7-8>   grxn0700
        <8-9>   grxn0800
        <9-10>  grxn0900
        <10-12> grxn1000
        <12-17> grxn1200
        <17->   grxn1728
      }{}

\begin{document} 
   \title{Reionisation time field reconstruction from 21-cm Maps: Investigating predictor coherence in WDM cosmology}

\author{
  Julien Hiegel\inst{1}$^,$\inst{2}$^,$\thanks{julien.hiegel@iphc.cnrs.fr}
  \and Dominique Aubert\inst{1}
  \and Émilie Thélie\inst{3}
  \and Rodrigo Ibata\inst{1}
  \and Nicolas Mai\inst{1}$^,$\inst{4}
}

\institute{
  Université de Strasbourg, CNRS UMR 7550, Observatoire Astronomique de Strasbourg, Strasbourg, France
  \and
  Institut Pluridisciplinaire Hubert Curien (IPHC), Université de Strasbourg, CNRS/IN2P3, 67037 Strasbourg, France
  \and
  Department of Astronomy, University of Texas at Austin, 2512 Speedway, Austin, TX 78712, USA
  \and
    Université Paris Cité, CNRS(/IN2P3), Astroparticule et Cosmologie, 75013 Paris, France
}

\date{Received ..., accepted ...}

\titlerunning{CNN Coherence for $\TR$ Reconstructions}
\authorrunning{Hiegel et al.}

\abstract  
    {The reionisation time field $\TR$ captures the entire history of cosmic reionisation by mapping the moment where each region of the Universe became ionised. Previous work (\cite{Hiegel_2023}) has shown that $\TR$ can be inferred from 21-cm observations, including SKA-like instrumental effect, using convolutional neural networks (CNNs). However, these CNN predictors are trained on specific reionisation models, raising critical concerns about their reliability when applied to observational data potentially differing from their training assumptions.}
    {This paper aims to propose and test a method to evaluate the coherence of our CNN predictors with respect to their input model, thereby enabling the validation or exclusion of underlying reionisation models based on their reconstruction behaviour.}
    {By setting the CDM model as reference input, we evaluate the coherence of $\TR$ reconstructions by comparing them across different redshifts for several prediction models as the statistics of $\TR$ reconstructions should be the same for every redshift of the input maps. Focusing on metrics such as the root-mean-square error (RMSE), the coherence fraction $q$, and the coefficient of determination $R_{\Delta}^2$, our study particularly investigates CNNs trained on cold and warm dark matter (WDM) models, with WDM particle masses of 2, 3, 5, and 7 keV.}
    {We find that the predictors trained on 5 and 7 keV WDM models exhibit high-level self-consistency similar to the CDM predictor, while the 2 keV predictor, and to a lesser extent the 3 keV predictor, display significant deviations across several metrics. These findings seem to demonstrate that CNN predictors retain sensitivity to differences in the underlying reionisation model and can be used to assess model compatibility with observations.}
    {Our results highlight the necessity of validating machine-learning predictors against their input models before applying them to real data. The method proposed here offers a pathway to more trustworthy applications of CNNs in the study of reionisation, and future work will aim to enhance model performance through improvements in CNN architectures and the adaptation to SKA-like observations.}
    
   \keywords{   Cosmology: large-scale structure of Universe, dark ages, reionisation, first stars, numerical simulation, Galaxies: formation, high-redshift, Deep Learning
               }

\maketitle

\section{Introduction}
\label{sec:Introduction}
Throughout its history, the universe has transitioned through distinct phases. Beginning with the production of the cosmic microwave background (CMB) at a redshift of $z\approx1100$, the universe entered the Dark Ages.
During this epoch, dark matter (DM) collapsed into haloes under the influence of gravity, eventually pulling baryonic matter into its potential wells to form the first structures such as stars and galaxies. The characteristic size of these early haloes was set by the free-streaming length $\lambda_{FS}$ of the dark matter particles: the smaller the particle mass $m_X$, the larger the suppression of small-scale structures (\cite{Blumenthal_1982}, \cite{Bode_2001}, \cite{Villasenor_2023}). Consequently, the nature of dark matter, particularly its particle mass, directly impacts the abundance and properties of the first haloes capable of forming galaxies.
Approximately 100 million years after the Big Bang, at $z\approx30$, the universe witnessed the birth of the first stars and galaxies within these dark matter haloes (\cite{Loeb_2001}, \cite{Wise_2019}), marking the onset of the Cosmic Dawn. These first light sources emitted X-ray and UV radiation, heating and reionising the intergalactic medium (IGM), mainly composed of neutral hydrogen and helium. Thus began the epoch of reionisation (EoR), which lasted hundreds of millions of years until $z\approx5-6$ (\cite{Kulkarni_2019}). The efficiency of early galaxy formation, tightly linked to the underlying dark matter distribution, played a critical role in setting the timing and topology of reionisation.
The exact mechanisms driving the ionisation process and its temporal and spatial variations remain elusive puzzles in the study of cosmic evolution. 

A key concept to understand the EoR is the distribution of neutral hydrogen HI. This element, while neutral, spontaneously releases a photon thanks to a spin flip transition in its fundamental state with the wavelength 21 cm. This signal is seen in absorption or emission with respect to the CMB, depending on the hydrogen content in the line of sight. On Earth, we can observe this redshifted signal in the range frequency of [90-200] MHz, corresponding to the redshift range [6, 15]. Radio-telescopes such as EDGES \cite{Bowman_2008}, PAPER (\cite{Parsons_2010}), LOFAR (\cite{Van_Haarlem_2013}), MWA (\cite{Tingay_2013}), HERA (\cite{DeBoer_2017}), NenuFAR (\cite{mertens_2021}), REACH (\cite{de_Lera_Acedo_2022}) and SKA (\cite{Koopmans_2015}) aim at observing this signal. Naturally, these observations come with instrumental effects and background contamination (\cite{Zaroubi_2012}, \cite{Munshi_2024}) that need to be studied to be ready for the incoming observations.
\subsection{Reionisation time}
Unlike HII maps or 21-cm observations which capture the ionisation state at a specific redshift, the reionisation time field $\TR$ (alternatively, the reionisation redshift $\ZR$, \cite{Trac_2008}, \cite{Battaglia_2013}, \cite{Deparis_2019}) captures the time (Gyr, alternatively the redshift z) at which each position $\Vec{r} =(x, y, z)$ in the Universe became reionised. 
By integrating the ionisation history over time, $\TR$ offers a cumulative and temporally rich view of the reionisation process. When projected as 2D maps, $\TR$ captures the spatial and temporal evolution of reionisation across cosmic structures, making it a powerful diagnostic tool offering the unique advantage of condensing the entire reionisation history of a region into a single field. Its study gives a direct and comprehensive understanding of when and where reionisation occurred, allowing for a clear identification of early sources, ionisation fronts, HII and HI regions, and late reionising voids (\cite{Thelie_2022}, \cite{Thelie_2023}). This also makes $\TR$ a valuable framework for linking the matter density of the Universe with the reionisation temporal evolution (\cite{Chardin_2019}), using e.g. AI. 

Although not directly observable, $\TR$ (or equivalently, $\ZR$) has traditionally been used to monitor the reionisation process in cosmological simulations such as, for example, $\cmfast$ (\cite{Mesinger_2011, Murray_2020}), EMMA (\cite{AUBERT_2015}), or AMBER (\cite{Trac_2022}). In addition, \cite{Hiegel_2023} demonstrated that convolutional neural networks (CNNs) can infer $\TR$ from 21-cm maps at a fixed redshift, offering a novel pathway to reconstruct the complete reionisation history from a single redshift observation. However, a key challenge lies in assessing the coherence of CNN-based predictors with the input data to which they are applied. Each predictor is inherently tied to a reionisation model embedded in its training set, which may not perfectly reflect the true physical conditions of the Universe. As a result, using such models for observational inference demands caution, particularly when the model's assumptions differ significantly from reality.

In this paper, we evaluate the coherence of our CNN predictors (\cite{Hiegel_2023}) with respect to the model used to generate their input maps. This allows us to assess whether a trained network is consistent with the underlying physical model that produced the mock data, and whether such a predictor can be reliably used in model selection or parameter inference. Although many parameters are satisfied, (e.g., ionisation efficiency $\zeta$, virial temperature $T_{\mathrm{vir}}$, source properties, etc.) can be adjusted to vary reionisation histories, we focus on dark matter property, considering warm dark matter (WDM). 

\textcolor{black}{WDM models were originally proposed to alleviate several small-scale challenges of the standard cold dark matter (CDM) framework, such as the cusp–core (\cite{Flores_1994}, \cite{Foidl_2023}), missing satellites (\cite{Klypin_1999}, \cite{Muller_2024}, \cite{Jung_2024}), and too-big-to-fail problems (\cite{Boylan-Kolchin_2011}, \cite{Kameli_2020}). However, recent studies have shown that many of these tensions can be significantly mitigated when baryonic processes are properly accounted for (e.g. \cite{Sales_2022}). Despite this, alternative dark matter scenarios, including WDM and mixed (WDM+CDM) models (\cite{Parimbelli_2021}, \cite{tadepalli2025warmdarkmattermeets}), remain viable and can lead to subtle but observable differences in structure formation. In this context, WDM provides a useful test case to assess the sensitivity of our method to deviations from the standard CDM scenario.}

Traditionally, the mass of the WDM particle is assumed to lie in the keV range. For instance, a WDM model with a particle mass of $m_X = 1$ keV allows the formation of galaxies with a minimum halo mass of approximately $10^{12}$ M$_\odot$ (\cite{Blumenthal_1982}), which implies that smaller galaxies require a heavier dark matter particle to form. However, the lower bound of $m_X$ remains an open question, and numerous studies have sought to constrain this parameter using various observational techniques. Recent analyses based on the Lyman-$\alpha$ forest from high-redshift quasar spectra observed with UVES and HIRES spectrographs have placed a lower limit of $m_X > 5.7$ keV (\cite{Irsic_2024}), a result corroborated by similar findings such as those in \cite{garciagallego2025constrainingmixeddarkmatter}. Further constraints have been obtained through alternative methodologies: for example, \cite{Nadler_2021} reported a limit of 9.7 keV from a combined analysis of strong gravitational lensing and Milky Way satellite galaxy counts, while \cite{Enzi_2021} derived a lower limit of 6.05 keV by combining Lyman-$\alpha$ forest data with strong lensing and satellite surveys.

In what follows, we explore how CNN predictors (\cite{Hiegel_2023}) trained on these WDM scenarios behave when confronted with mock observations based on CDM models. This analysis not only tests the robustness of CNN inference, but also sheds light on the sensitivity of $\TR$ reconstruction to the underlying dark matter physics. 
Section 2 introduces the models used in our study and their key differences. Section 3 presents the $\TR$ reconstruction and evaluates CNN performance when the input matches the training model. In Section 4, we evaluate the self-consistency of CNN reconstructions obtained with CDM input. Eventually, section 5 offers a brief discussion, followed by the conclusion in Section 6.
\section{Data set - dark matter models}
\label{sec:Models}
To obtain our data set, we employed the $\cmfast$ simulation code (\cite{Mesinger_2011}, \cite{Murray_2020}). We generated coeval simulation cubes with a spatial extent of 256 cMpch$^{-1}$ at a resolution of 1cMpch$^{-1}$/pixel. These simulations were conducted within the framework of a $\Lambda$CDM cosmology, characterised by the cosmological parameters ($\Omega_m,\Omega_b,\Omega_{\Lambda},h,\sigma_8,n_s$) = (0.31, 0.05, 0.69, 0.68, 0.81, 0.97), consistent with the findings reported by Planck 2018. We set the parameter $\zeta$ (see \cite{Greig_2015}) to 30 which sets the ionising efficiency of high-z galaxies: the larger the value, the sooner the reionisation ends. In addition, we use the default value of log$_{10}T_{\mathrm{vir}}$=4.7. This parameter sets the minimal virial temperature for halos to host star formation (\cite{Barkana_2001}, \cite{Gillet_2016_phd}).
In addition, we considered four WDM models with the same parameter values as those for the CDM model. These WDM models incorporate thermal relic particles with masses 2 keV, 3 keV, 5 keV, and 7 keV, respectively. All five models experienced reionisation that concluded at approximately z$\approx$5. For the 2 and 3 keV models, the value of $\zeta$ was adjusted to 32 to ensure consistent reionisation histories.
The 2 and 3 keV models, yet extensively ruled out in the literature, will serve as references to test our ability to exclude models that significantly deviate from the CDM scenario.

\begin{figure}
    \centering
    \includegraphics[width=0.5\textwidth]{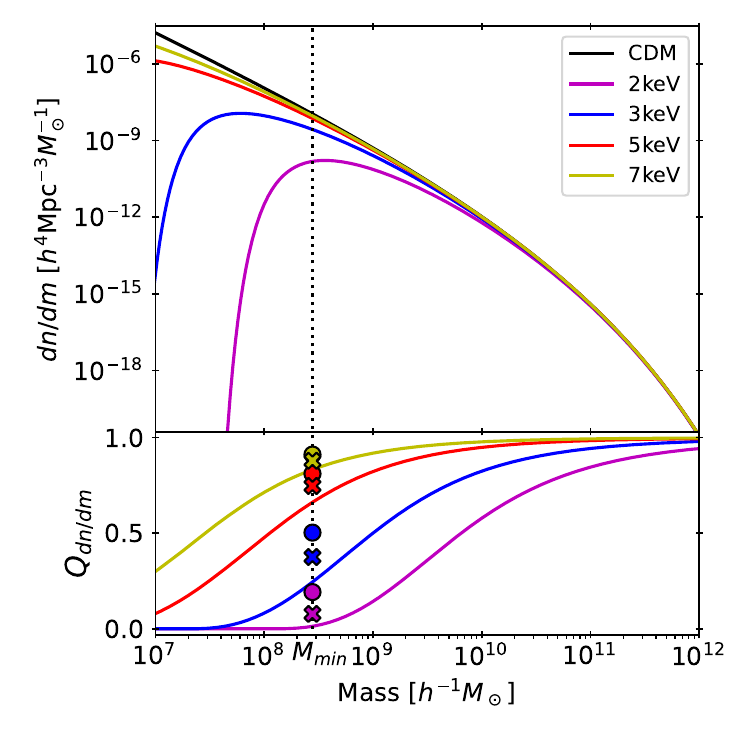}
    \caption{Theoretical halo mass function for the five models used in this paper at $z=8$. The vertical dotted line shows the minimal halo mass required to host star formation, computed from $T_{\mathrm{vir}}$. The bottom panel shows the fraction $n_{\mathrm{wdm}_i}(m)/n_{\mathrm{cdm}}(m)$, where $n(m)=dn/dm$ and $\mathrm{wdm}_i$ is the WDM model associated with the relevant colour.  Cross markers display the fraction $n_{\mathrm{wdm}_i}(>M_{\mathrm{min}})/n_{\mathrm{cdm}}(>M_{\mathrm{min}})$ where $n(>M_{\mathrm{min}})$ is the number of halo with mass greater than $M_{\mathrm{min}}$. Plus markers display the fraction $M_{\mathrm{wdm}_i}(>M_{\mathrm{min}})/M_{\mathrm{cdm}}(>M_{\mathrm{min}})$ where $M(>M_{\mathrm{min}})$ is the total mass within haloes of mass greater than $M_{\mathrm{min}}$ for each model.}
    \label{fig:hmf}
\end{figure}

\subsection{Halo mass function}
Even though the reionisation scenario is similar, the five models are expected to exhibit differences. For instance, the matter density power spectrum $P_m(k)$ for WDM models is expected to show a cutoff at large spatial frequencies k, indicating fewer structures at small scales. This phenomenon directly impacts the halo mass function (HMF) as shown in Fig. \ref{fig:hmf}. The HMF was computed using a standard method as proposed in \cite{Barkana_2001} (also see \cite{Sheth_2001}, \cite{Maggiore_2010}, \cite{Benson_2012} and \cite{Schneider_2013}):
\begin{equation} \label{eq:hmf}   
    \frac{dn}{dM}= \frac{\rho_m}{M} \left(\frac{-dln\sigma}{dM}\right) \nu f(\nu).
\end{equation}
Here, $\rho_m$ represents the mean matter density of the universe at the current day (\cite{Eke_1996}, equation 2.1) and $\sigma$ is the fractional root variance of the mass density field computed from the matter power spectrum $P_m(k)$ at a specific redshift. 
In this study, the CDM power spectrum was obtained using CAMB\footnote{https://camb.readthedocs.io/en/latest/} (Code for Anisotropies in the Microwave Background, \cite{Lewis_2000}). A correction is applied to obtain WDM ones by multiplying the CDM power spectrum with a transfer function as described by (\cite{Viel_2005}):
\begin{equation} 
\label{eq:pwdm}   
    P_{\mathrm{wdm}}(k) = P_{\mathrm{cdm}}(k)\times T(k)^2,
\end{equation}
where, $T(k)= [1 + (\alpha k)^{2\eta}]^{-5/\eta}$, with $\eta = 1.12$  and $\alpha$ is the scale of the break, depending of the WDM mass (refer to \cite{Viel_2005}, section II equation (6)).
The parameter $\nu$ in Eq. \ref{eq:hmf} is defined as $\nu=\delta_c/\sigma$, where $\delta_c$ is the critical collapse over-density assumed to be 1.686 (\cite{Eke_1996} and \cite{Jenkins_2001}, section 4). Additionally, we have:
\begin{equation} \label{eq:nufnu}   
    \nu f(\nu)=\nu\sqrt{\frac{2}{\pi}} exp\left(-\frac{\nu^2}{2}\right).
\end{equation}

The minimal mass of star-forming haloes $M_{\mathrm{min}}$ (vertical dotted line) is computed from $T_{\mathrm{vir}}$ and is defined as follows (\cite{Greig_2015}, see also \cite{Barkana_2001b}):
\begin{equation} 
\label{eq:Mmin}   
    M_{\mathrm{min}} = 10^8 h^{-1} 
    \left( \frac{\mu}{0.6} \right)^{-3/2} 
    f(\Omega) 
    \left( \frac{T_{\mathrm{vir}}}{1.98\times 10^4}\right)^{3/2} 
    \left( \frac{1+z}{10}\right)^{-3/2} M_\odot
\end{equation}
 where $\mu$ is the  mean molecular weight and $f(\Omega)$ is a function depending on the contribution of matter and vacuum densities ($\Omega_m$ and $\Omega_{\Lambda}$, see \cite{Greig_2015}, Eq. 3). In our case, the value of $M_{\mathrm{min}}$ at $z=8$ is $2.76\times10^8 M_{\odot}$. 
 The HMF for the WDM models closely aligns with the CDM at mass $M=10^{12}M_{\odot}$, ranging from 94, 98, 99, 100$\%$ of CDM abundance at $10^{12}M_{\odot}$ and decreasing to 1, 24, 66, 83$\%$ at $M_{\mathrm{min}}$ for masses 2, 3, 5, 7 keV, respectively. This was expected given the sharp decline in the number of small structures forming in WDM cosmology. In the bottom panel, the ratio $\mathrm{wdm}_{i}/\mathrm{cdm}$ mass functions is depicted, with $\mathrm{wdm}_{i}$ denoting the considered WDM model. The coloured curves display the HMF fraction at each mass. 
The cross and plus markers represent the quantities $Q_n$ and $Q_M$, defined as the ratios $Q_n = n_{\mathrm{WDM}}/n_{\mathrm{CDM}}$ and $Q_M = M_{\mathrm{WDM}}/M_{\mathrm{CDM}}$, respectively, where
\begin{equation}
\label{eq:ntot}
n(>M_{\mathrm{min}})=\int_{M_{\mathrm{min}}} \mathrm{d}m\,\frac{\mathrm{d}n}{\mathrm{d}m},
\end{equation}
and
\begin{equation}
\label{eq:mtot}
M(>M_{\mathrm{min}})=\int_{M_{\mathrm{min}}} \mathrm{d}m\,\frac{\mathrm{d}n}{\mathrm{d}m}\,m,
\end{equation}
respectively. Here, $n(>M_{\mathrm{min}})$ denotes the total number of haloes and $M(>M_{\mathrm{min}})$ the total mass of haloes contributing to reionisation with masses larger than $M_{\mathrm{min}}$. For WDM particle masses of 2, 3, 5 and 7 keV, the corresponding pairs of values $[Q_n, Q_M]$ are $[0.08, 0.19]$, $[0.38, 0.50]$, $[0.75, 0.81]$ and $[0.88, 0.91]$, respectively.
The number of halo contributing to reionisation differs significantly for the 2, 3 and 5 keV models. Furthermore, a smaller halo emits fewer ionising photons than a more massive one, and therefore, even though smaller haloes contributing to reionisation are more numerous, the heavier haloes have a substantial impact in terms of contributing mass and thus photons. From this preliminary study of our models, the 7 keV model is expected to display results similar to CDM, while the 2 and 3 keV models should present larger differences and be easily distinguishable. The 5 keV model falls in between and is expected to show an intermediate behaviour.

\begin{figure*}
    \centering
    \includegraphics[width=1\textwidth]{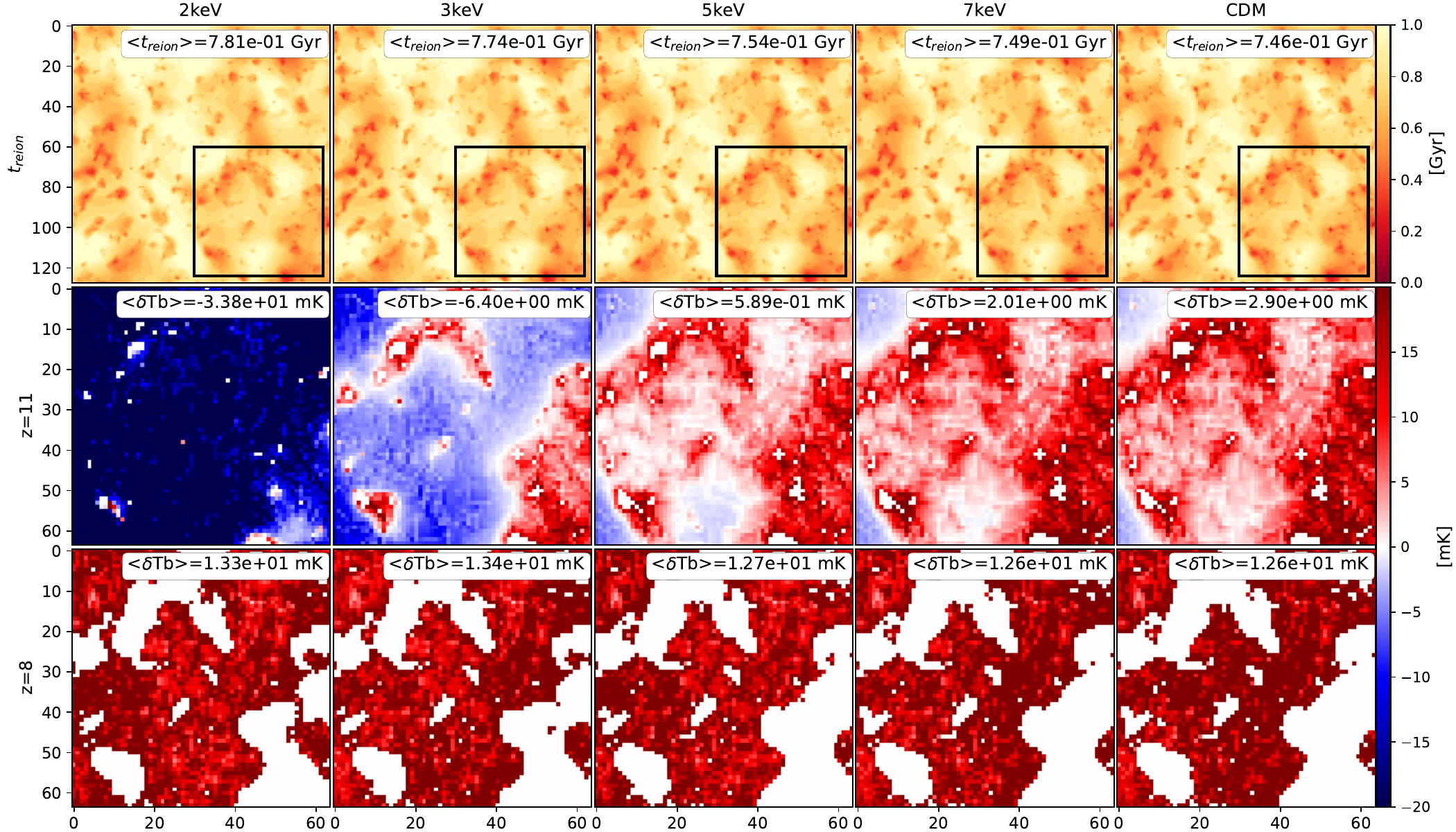}
    \caption{Illustration of the five DM models and their respective fields. \textcolor{black}{The upper row depicts the reionisation time field, while the two rows below show 21-cm maps at redshifts 11 and 8 extracted from the region indicated by the black square in the $\TR$ maps.} Each column represents a specific DM model, arranged from left to right as follows: 2 keV, 3 KeV, 5 keV, 7 keV, and CDM. These maps have been produced using $\cmfast$ simulation code. The unit of x and y-axis is [$\mpc$].}
    \label{fig:treion_tb_maps}
\end{figure*}
\subsection{Data set - maps}
Figure \ref{fig:treion_tb_maps} illustrates representative 2D maps from our data set. The top row showcases the reionisation time field, where first inspection might not reveal differences among models, despite variations in mean reionisation times. Moving to the middle row, the 21-cm maps at z=11 illustrate a lower mean temperature brightness $\delta T_b$ for lighter WDM particles: the distribution of temperature is different because the reionisation timing is different as seen in the mean reionisation time values on the top row maps. There are also missing regions as the WDM particle mass decreases: for example, in the CDM model, hot regions (in red) are present around (x, y) = (20, 10) cMpch$^{-1}$, these regions get smaller for the 7 keV model, start to disappear for the 5 keV model and have vanished for the 3 keV and 2 keV models. Eventually, the bottom row depicts 21-cm maps at z=8, showing that the temperature brightness for WDM models has now a similar shape and mean value as CDM: a result aligned with our expectation of reionisation scenarios ending around the same time. However, some differences can be observed, for example, the reionised region (in white) at (x, y) = (55, 40) cMpc$^{-1}$, which becomes tighter as the mass of the WDM particles decreases and disappears for model 2 keV. 
\subsection{Neutral fraction - $Q_{\mathrm{HI}}$}
\begin{figure}
    \centering
    \includegraphics[width=0.5\textwidth]{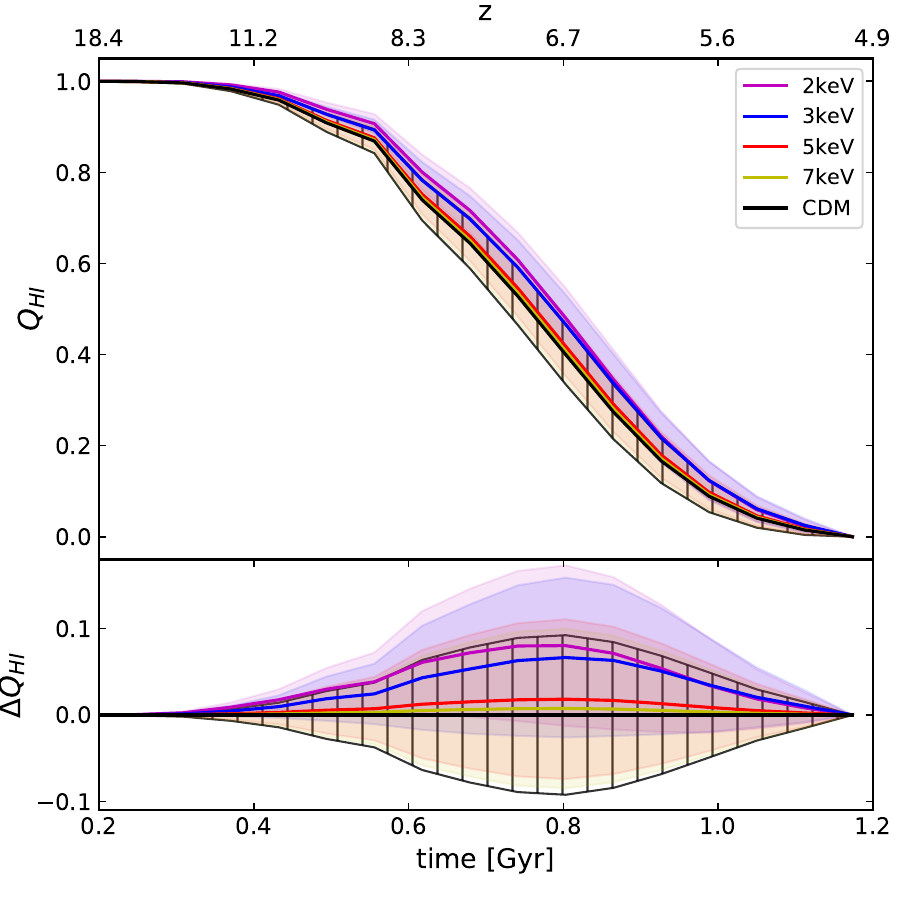}
    \caption{Volume fraction of neutral hydrogen $Q_{\mathrm{HI}}$ for the five DM models. Time goes from left to right. The upper panel describes the mean reionisation history. On the left, all hydrogen is neutral (=1) and on the right, all hydrogen is ionised (=0). The lower panel displays the difference $Q_{\mathrm{HI,WDM}}$-$Q_{\mathrm{HI,CDM}}$ where  WDM is the corresponding model according to the colour. Shaded areas stand for the standard deviation for each model.}
    \label{fig:QHI_true}
\end{figure}

As shown in Fig. \ref{fig:QHI_true}, the volume fraction of neutral hydrogen only slightly differs between models: up to an 8$\%$ difference at around 800 Myr (z$\approx$6.7) for the model 2 keV compared to the CDM that decreases to less than 2$\%$ and 1$\%$ for model 5 keV and 7 keV respectively at the same cosmological time. It is crucial to note that the WDM particle mass has a similar role to the $\zeta$ parameter, altering the timing of reionisation. While the former mainly affects the quantity and mass of structures, the latter influences the ionising efficiency of each structure. Therefore, as the mass decreases, it becomes necessary to enhance $\zeta$ to achieve equivalent reionisation timing. However, masses of 5 and 7 keV appear sufficiently heavy to maintain a reionisation history that aligns with the CDM scenario.
\subsection{Reionisation time power spectrum}
\begin{figure}
    \centering
    \includegraphics[width=0.5\textwidth]{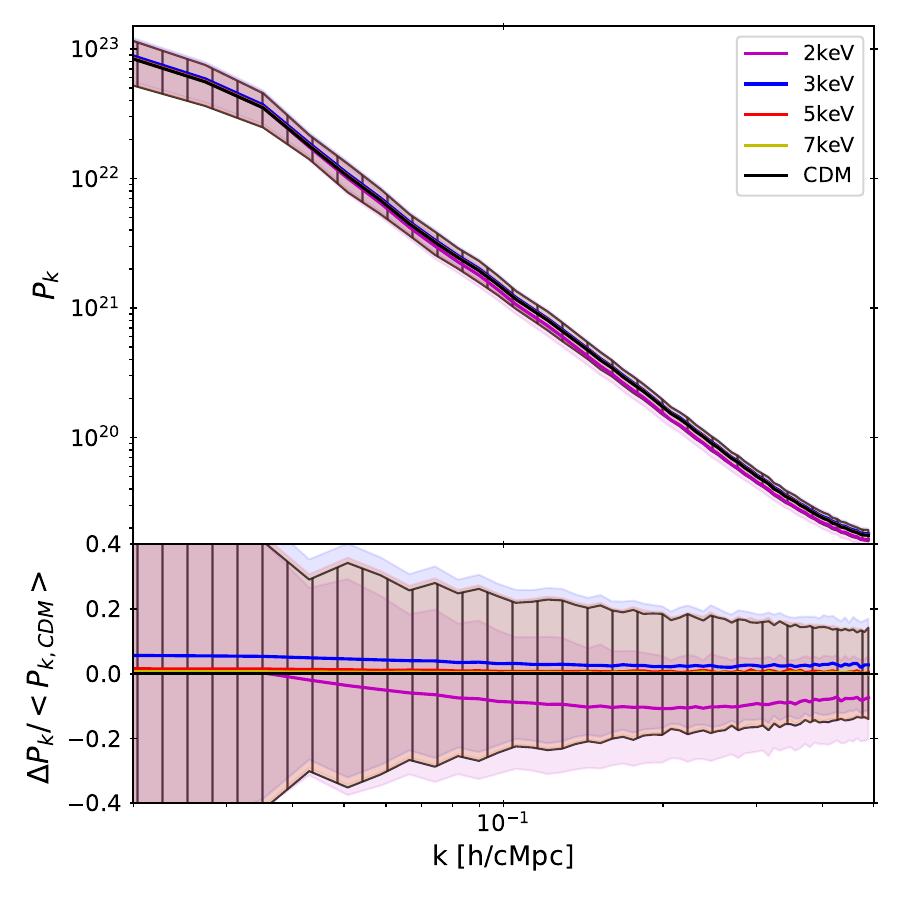}
    \caption{Power Spectrum $P_{k}$ computed from $\TR$ for the five DM models. The upper panel describes the power spectrum, both mean (solid lines) and standard deviation (shaded areas). The large-scale structures are represented on the left side of the figure, while the small scale structures are depicted on the right (respectively small and large spatial frequencies). The lower panel displays the relative difference between $P_{k,\mathrm{WDM}}$ and $P_{k,\mathrm{CDM}}$ where  WDM is the corresponding model according to the colour.}
    \label{fig:Pk_true}
\end{figure}

Figure \ref{fig:Pk_true} presents the $\TR$ power spectrum $P_k$ for each model in the top panel and the relative difference between the WDM models and the CDM power spectra in the bottom panel. From this metric, the five models are consistent and only a 10$\%$ difference at max is found for the 2 keV model. Also note that the reionisation time power spectrum for this model has a different shape than the other models: this mass starts to become too small and implies fewer structures at almost every spatial frequency. Indeed, as the WDM particle mass decreases, fewer halos form, leading to a reduction in the number of galaxies that emerge. In a universe with fewer galaxies, there are consequently fewer reionisation seeds, resulting in fewer minima in the $\TR$ field and then fewer observable structures.
\subsection{Isocontour length - HI and HII regions}
\begin{figure}
    \centering
    \includegraphics[width=0.5\textwidth]{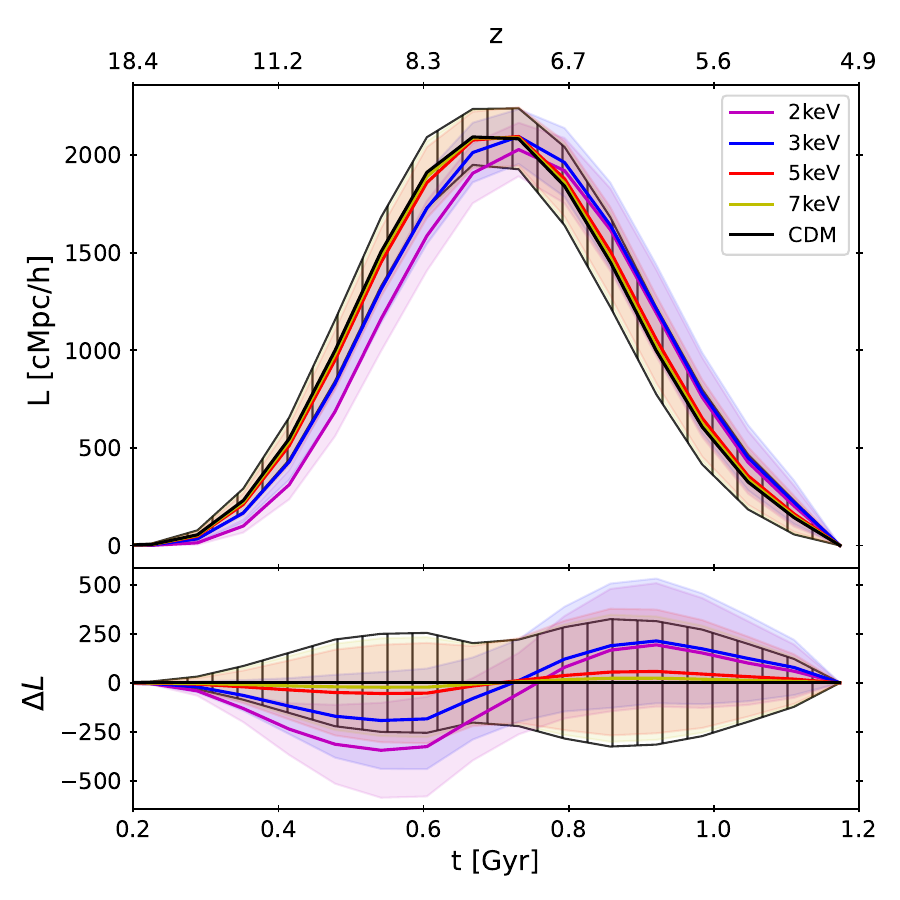}
    \caption{Comparison of the total length of isocontours between the five DM models. Time goes from left to right. The upper panel shows the total isocontour length per map ($128^2$ $\mathrm{cMpc}^2/h^2$) with respect to to the cosmological time. The lower panel displays the difference $L_{\mathrm{WDM}}-L_{\mathrm{CDM}}$ where WDM is the corresponding model according to the colour. Shaded areas are the standard deviation of each model.}
    \label{fig:L_true}
\end{figure}

We compute the $\TR$ isocontours for each model to obtain the total isocontour length found for each map for a given reionisation time. We take then the average across maps for each model and the result is shown in Fig. \ref{fig:L_true}. This distribution displays the cumulative perimeter of hydrogen bubbles at a given time. While the isocontours illustrate the interface between ionised and neutral regions at a specific time, they present either the boundaries of ionised regions or the persisting neutral regions during early and late times, respectively. 
Then, at early times (<0.4 Gyr) the isocontours mainly depict bubbles of ionised gas spreading around seeds of reionisation. At late times (>1 Gyr), the isocontours rather depict the remaining neutral gas bubbles\footnote{Note the use of "mainly" and "rather": it is a trend, but there might be exceptions: Think about a late source of reionisation appearing in a neutral region at late times.}. From this figure, WDM scenarios have smaller ionised bubbles at early times (<0.6 Gyr): as the WDM particle mass decreases, the simulation forms lighter and fewer structures, enabling a slower reionisation process. For example according to this metric, at 0.41 Gyr: there are 43, 21, 6 and 3$\%$ length deficits for the models 2, 3, 5 and 7 keV respectively. On the opposite, at late times (>0.8 Gyr), WDM models have larger remaining neutral regions: as the reionisation process "started slower", more regions have not been reionised yet and the end of the EoR is thus slightly delayed. At 1 Gyr, there is 25, 28, 7, 3$\%$ more length for the neutral regions in comparisons to the CDM for the models 2, 3, 5, 7 keV.
\subsection{Minima - reionisation sources}
\label{sec:minima_norm}
\begin{figure}
    \centering
    \includegraphics[width=0.5\textwidth]{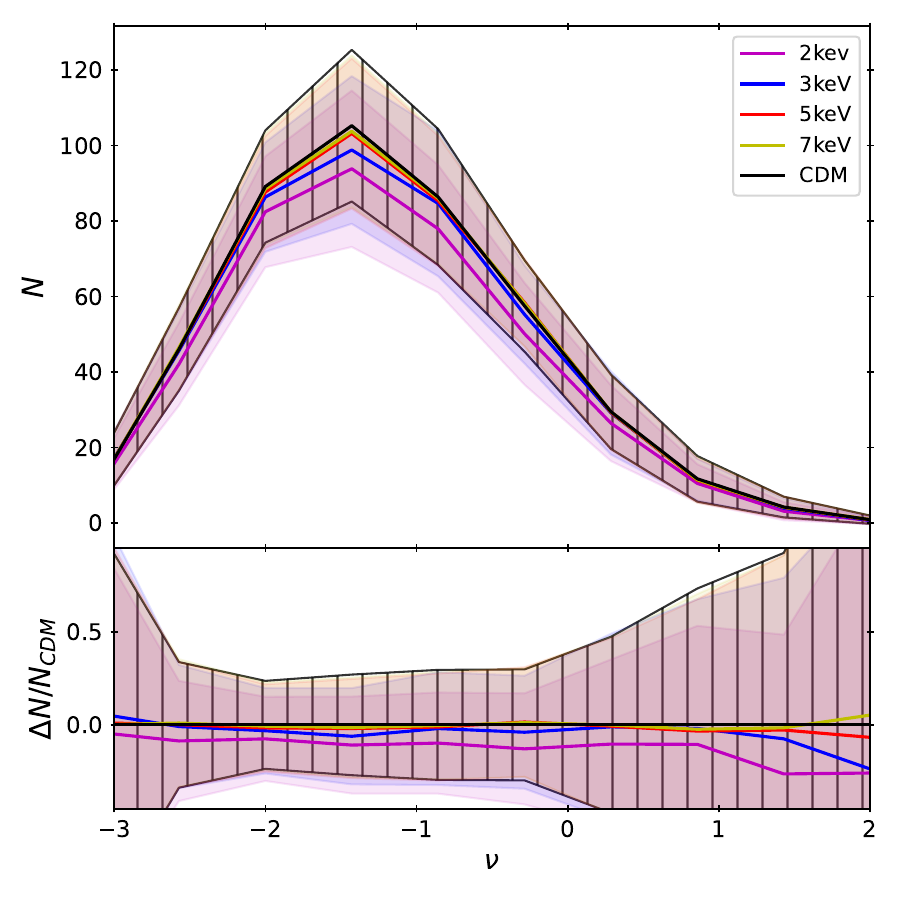}
    \caption{Minima number count for the five DM models depicting the minima of $\TR$. Time goes from left to right and the axis is standardised where 0 is the mean reionisation time for each model. The upper panel shows the mean (solid lines) and standard deviation (shaded areas) of the minima number count. The lower panel displays the relative difference $N_{\mathrm{WDM}}-N_{\mathrm{CDM}}$ divided by $N_{\mathrm{CDM}}$ where WDM is the corresponding model according to the colour.}
    \label{fig:N_true}
\end{figure}

The last metric used to differentiate models is shown in Fig. \ref{fig:N_true}. It displays the number of minima with respect to time. Here, the time has been standardised by subtracting the mean and dividing by the standard deviation, such as follows:
\begin{equation}
   \nu_{\TR} = \frac{\TR - <\TR>}{\sigma_{\TR}}
\end{equation}
where, for one map,  $\TR$ is the true pixel values of the map, <$\TR$> and $\sigma_{\TR}$ are the mean and the standard deviation of the reionisation time field data set, respectively and $\nu_{\TR}$ is the normalised map. Hence, $\nu_{\TR}=0$ represents the mean reionisation time (see the mean values in Fig. \ref{fig:treion_tb_maps}). 
To identify local minima within the reionisation time fields, we developed a custom detection algorithm rather than relying on the DisPerSE code (Discrete Persistent Structures Extractor; \cite{Sousbie_2011}, \cite{Sousbie_2011b}) as previously in \cite{Thelie_2022} and \cite{Hiegel_2023}. The method performs an exhaustive, pixel-by-pixel search for local minima and plateau structures. For each pixel, its value is compared to its immediate neighbours in a cross-shaped connectivity. If the pixel is strictly lower than its neighbours, it is flagged as a local minimum. In the case of plateaus (connected regions of equal intensity) a flood-fill procedure is applied to explore the entire plateau and determine whether the surrounding values are systematically higher or lower. A plateau is classified as a minimum if no surrounding pixel has a smaller value. To ensure robust detection, the algorithm also handles boundary conditions explicitly, allowing minima located at the edges of the field to be properly identified. For each detected minimum or minimum plateau, a single representative coordinate is assigned, corresponding to the barycentre of the plateau region. This approach guarantees that each local minimum is counted exactly once, avoids multiple detections on flat structures, and does not rely on derivatives or threshold.

The result is shown in the figure \ref{fig:N_true}. Since it focuses on $\TR$ minima, the distribution is shifted towards the negative values of the normalised time $\nu$ because the minima are the seeds of reionisation, likely the sources, that mostly appeared before the mean reionisation time. The main top panel shows the minima number, while the bottom panel shows the relative differences between WDM models and the CDM. For this metric, the five models are fairly close to one another, following a trend in which the heavier the mass, the more minima are found. The 5 and 7 keV model curves are actually quite close to the CDM ones, while we observe a larger discrepancy for the 2 and 3 keV models. This behaviour was anticipated, as it is typical of WDM models.

 \section{Reionisation time reconstruction}
\label{sec:TR_reconstruction}

This study takes root in a previous one dedicated to the $\TR$ reconstruction from mock 21-cm maps. In this former paper (\cite{Hiegel_2023}), we used a CNN (U-net architecture, \cite{ronneberger2015}) to infer $\TR$ from 21-cm maps taken at a given redshift. We found that the reionisation time can be inferred from the pure 21-cm signal, and in such reconstructions, the large-scale structures are well-defined even though the small scales get smoothed. Also, a trend has been found where the observation redshift range $\in$ [8,12] is optimal to infer the desired field for a similar CDM reionisation scenario such as described in Sec. \ref{sec:Models}. Not only the reionisation time information is encoded into the 21-cm signal, but the predictor is capable of inferring when a pixel in the 21-cm map will reionise in the future of the redshift at which the 21-cm observation is done, or when it has reionised in its past. Eventually, incorporating SKA-like instrumental effects into the mock data complicates the inference of the reionisation time, resulting in further smoothing of the predicted maps (also see \cite{Bianco_2021}, \cite{chen2023stabilitydeeplearning21cm}, \cite{Bianco_2024}, \cite{Kennedy_2024} and \cite{bianco2024deeplearningapproachidentification} for more insights about instrumental systematics and foreground contamination). To estimate the overall performance of our CNN, we use the coefficient of determination $R^2$, defined as follows:
\begin{equation}
\label{eq:R2}   
    R^2=1-\frac{\sum (X-X_0)^2}{\sum(X_0-<X_0>)^2}.
\end{equation}
Here, X stands for the predicted maps and $X_0$ for the ground truth. The summation is performed over each pixel of the images in the validation set.  
In practice, the $R^2$ value ranges from 0 to 1, where 1 indicates a perfect correlation, typically X = $X_0$. Nevertheless, $R^2$ can be negative for various reasons, such as when X and $X_0$ are uncorrelated or when the predictor\footnote{A predictor here is a CNN trained with a given physical model and a given redshift, see section \ref{sec:training_predictors}.} performs poorly compared to a simplistic model aiming to predict the mean value. In \cite{Hiegel_2023}, $R^2$ values of [0.88, 0.85] were found for $z=11$ and 8 respectively. This suggests a strong correlation between the predicted $\TR$ maps and the ground truth ($\TR$ maps obtained from the simulation), indicating the fair performance of the predictor in capturing the desired features.

\textcolor{black}{Furthermore, we estimated the epistemic uncertainty associated with the CNN reconstruction using Monte Carlo dropout, as detailed in Appendix~\ref{app:D:sigma_model}.
For a given observable X (see Sec.~\ref{sec:4_WDM}), the corresponding model uncertainty is denoted $\sigma_X$ and is computed from multiple stochastic forward passes of the network. This quantity characterises the confidence of the CNN in its predictions.
We compared $\sigma_X$ to the standard deviation of the same observable measured across independent reconstructions of the test data set, denoted $\mathrm{std}_X$. The latter reflects the intrinsic variability of the reconstructed observable between predictions. Its behaviour for the different observables is illustrated in the figures presented in Secs.~\ref{subsec:QHI}, \ref{subsec:Pk}, \ref{subsec:L}, and \ref{subsec:N}.
Finally, we evaluated the ratio $\Delta\sigma=\sigma_X$/$\mathrm{std}_X$ in order to assess whether the uncertainty introduced by the reconstruction model remains subdominant with respect to the intrinsic scatter of the data. Values of $\Delta\sigma \ll1$ indicate that the epistemic uncertainty of the CNN is small compared to the variability of the observable across the test set.}

\subsection{Predictors - trained CNNs for reionisation reconstruction}
\label{sec:training_predictors}
In this paper, we used the five previously presented WDM models to check the feasibility of distinguishing models with CNNs. For each model and each redshift (see Sect. \ref{sec:Models}), a unique CNN predictor is considered. In the following, we refer to a predictor as a CNN trained on 21-cm maps generated from a specific dark matter model (either WDM or CDM) at a given redshift ($z = 8$ or $z = 11$). For example, we refer to the CNN trained with 3 keV WDM maps at $z = 11$ as the "3 keV predictor at $z = 11$", and similarly for other configurations. The objective of a predictor is to reconstruct the $\TR$ field from a single 21-cm map at a given redshift. 

For each predictor, the learning stage was processed using 35 000 images, including 31 500 images for the training data set and 3 500 images for the validation data set. The output images are normalised as described in Section~\ref{sec:minima_norm}, using the mean and standard deviation computed over the entire dataset, but the input images are normalised with the same expression using the mean and standard deviation of each individual map.
We use the same U-net architecture as in \cite{Hiegel_2023} consisting of several convolution+maxpooling layers (encoder part) and convolution+upsampling layers (decoder part). The loss function used is the mean square error (MSE) and the optimizer is \textit{adam}. Finally, we stopped the training phase at the epoch 100 for each predictor, where the performance converged a few epochs before.

As a preliminary validation step, it is important to verify that all predictors provide similar performances and consistent reconstructions when applied to input data generated from their own underlying model. The corresponding raw predictions and $R^2$ scores are presented in appendix \ref{app:A} and show that no predictor over-performs the others.

\begin{figure*}[h!]
    \centering
    \begin{tikzpicture}[
        node distance=1.2cm and 1.6cm,
        box/.style={draw, rounded corners, minimum width=2cm, minimum height=1.1cm, align=center},
        arrow/.style={-{Latex}, thick},
    ]
    \node[] (cdmMid) at (0, 0) {}; 
    \node[draw, rounded corners, minimum width=3cm, minimum height=3cm, above=0.1cm of cdmMid,
        path picture={
          \path[clip, rounded corners] 
            (path picture bounding box.south west) rectangle (path picture bounding box.north east);
          \node[anchor=center, inner sep=0pt] at (path picture bounding box.center) 
            {\includegraphics[width=3cm,height=3cm]{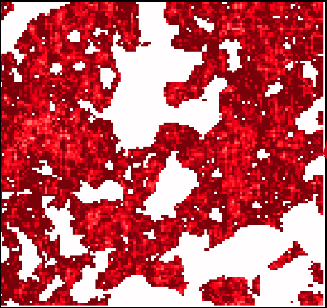}};
        }, align=center, text=black, font=\bfseries ] (cdm8) {\phantom{Input CDM $z=8$}};
    \node[draw, rounded corners, minimum width=3cm, minimum height=3cm, below=0.1cm of cdmMid,
        path picture={\path[clip, rounded corners] (path picture bounding box.south west) rectangle (path picture bounding box.north east);
      \node[anchor=center, inner sep=0pt] at (path picture bounding box.center) 
        {\includegraphics[width=3cm,height=3cm]{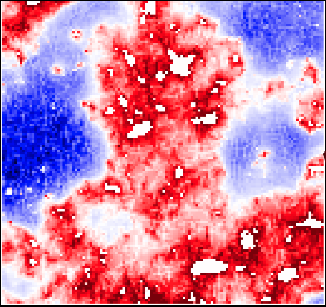}};
    }, align=center, text=black, font=\bfseries ] (cdm11) {\phantom{Input CDM $z=11$}};
    
    \node[box,minimum width = 3.0cm, right=1.5cm of cdm8, yshift=0.8cm] (predCDM8) {Pred CDM $z=8$}; 
    \node[box,minimum width = 3.0cm, right=1.5cm of cdm8, yshift=-0.8cm] (predCDM11) {Pred CDM $z=11$};
    
    \node[box,minimum width = 3.0cm, right=1.5cm of cdm11, yshift=0.8cm] (predWDM8) {Pred WDM $z=8$};
    \node[box,minimum width = 3.0cm, right=1.5cm of cdm11, yshift=-0.8cm] (predWDM11) {Pred WDM $z=11$};
    
    \node[draw, rounded corners, minimum width=1.5cm, minimum height=1.5cm, right=of predCDM8,
        path picture={
          \path[clip, rounded corners] 
            (path picture bounding box.south west) rectangle (path picture bounding box.north east);
          \node[anchor=center, inner sep=0pt] at (path picture bounding box.center) 
            {\includegraphics[width=1.5cm,height=1.5cm]{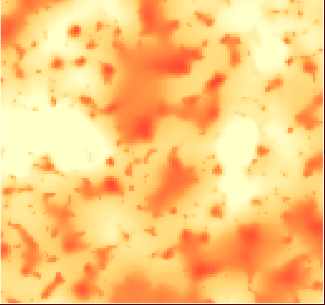}};
        }, align=center, text=black, font=\bfseries ] (trCDM8) {};
    \node[draw, rounded corners, minimum width=1.5cm, minimum height=1.5cm, right=of predCDM11,
        path picture={
          \path[clip, rounded corners] 
            (path picture bounding box.south west) rectangle (path picture bounding box.north east);
          \node[anchor=center, inner sep=0pt] at (path picture bounding box.center) 
            {\includegraphics[width=1.5cm,height=1.5cm]{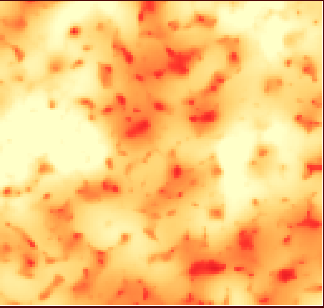}};
        }, align=center, text=black, font=\bfseries ] (trCDM11) {};
    
    \node[draw, rounded corners, minimum width=1.5cm, minimum height=1.5cm, right=of predWDM8,
        path picture={
          \path[clip, rounded corners] 
            (path picture bounding box.south west) rectangle (path picture bounding box.north east);
          \node[anchor=center, inner sep=0pt] at (path picture bounding box.center) 
            {\includegraphics[width=1.5cm,height=1.5cm]{treionz8.png}};
        }, align=center, text=black, font=\bfseries ] (trWDM8) {};
    \node[draw, rounded corners, minimum width=1.5cm, minimum height=1.5cm, right=of predWDM11,
        path picture={
          \path[clip, rounded corners] 
            (path picture bounding box.south west) rectangle (path picture bounding box.north east);
          \node[anchor=center, inner sep=0pt] at (path picture bounding box.center) 
            {\includegraphics[width=1.5cm,height=1.5cm]{treionz11.png}};
        }, align=center, text=black, font=\bfseries ] (trWDM11) {};
    
    \node[box, right=of trCDM8] (qhiCDM8) {$Q_{\mathrm{HI}}^{z=8}$(CDM)};
    \node[box, right=of trCDM11] (qhiCDM11) {$Q_{\mathrm{HI}}^{z=11}$(CDM)};
    \node[box, right=of trWDM8] (qhiWDM8) {$Q_{\mathrm{HI}}^{z=8}$(WDM)};
    \node[box, right=of trWDM11] (qhiWDM11) {$Q_{\mathrm{HI}}^{z=11}$(WDM)};
    
    \node[box, right=13cm of cdm8] (deltaCDM) {$\Delta Q_{\text{H\,I}}$ (CDM)};
    \node[box, right=13cm of cdm11] (deltaWDM) {$\Delta Q_{\text{H\,I}}$ (WDM)};
    
    \draw[arrow, bend left = 20] (cdm8.east) to (predCDM8.west);
    \draw[arrow, bend right = 30] (cdm8.east) to (predWDM8.west);
    \draw[arrow, bend left = 30] (cdm11.east) to (predCDM11.west);
    \draw[arrow, bend right = 20] (cdm11.east) to (predWDM11.west);
    
    \draw[arrow] (predCDM8) -- (trCDM8);
    \draw[arrow] (predWDM8) -- (trWDM8);
    \draw[arrow] (predCDM11) -- (trCDM11);
    \draw[arrow] (predWDM11) -- (trWDM11);
    
    \draw[arrow] (trCDM8) -- (qhiCDM8);
    \draw[arrow] (trCDM11) -- (qhiCDM11);
    \draw[arrow] (trWDM8) -- (qhiWDM8);
    \draw[arrow] (trWDM11) -- (qhiWDM11);
    
    \draw[arrow, bend left=15] (qhiCDM8.east) to ([yshift=2]deltaCDM.west);
    \draw[arrow, bend right=15] (qhiCDM11.east) to ([yshift=-2]deltaCDM.west);
    
    \draw[arrow, bend left=15] (qhiWDM8.east) to ([yshift=2]deltaWDM.west);
    \draw[arrow, bend right=15] (qhiWDM11.east) to ([yshift=-2]deltaWDM.west);
    
    \draw[thick, <->] (deltaCDM) -- (deltaWDM);
    
    \end{tikzpicture}    
    \caption{Pipeline used to obtain the coherence of our predictors with respect to CDM inputs (on the left, for redshift 8 on the top and redshift 11 on the bottom). For the example, only two models is represented (CDM and WDM) and we consider the statistic $Q_{\mathrm{HI}}$. First, each predictor is fed with the 21-cm CDM map at the corresponding redshift to obtain $\TR$ reconstruction. Afterward, we compute $Q_{\mathrm{HI}}$ for each reconstruction and finally substract $Q_{\mathrm{HI}}^{z=8}$ from $Q_{\mathrm{HI}}^{z=11}$ to obtain $\Delta Q_{\mathrm{HI}}$ for each model. In the scenario, $\Delta Q_{\mathrm{HI}}$(CDM) is the self-consistency reference, since the input model aligns with the predictor underlying model and we compare it to the WDM self-consistency (vertical arrow).}
    \label{fig:pipelineWDM}
\end{figure*}
\begin{figure*}
    \centering
    \includegraphics[width=1\textwidth]{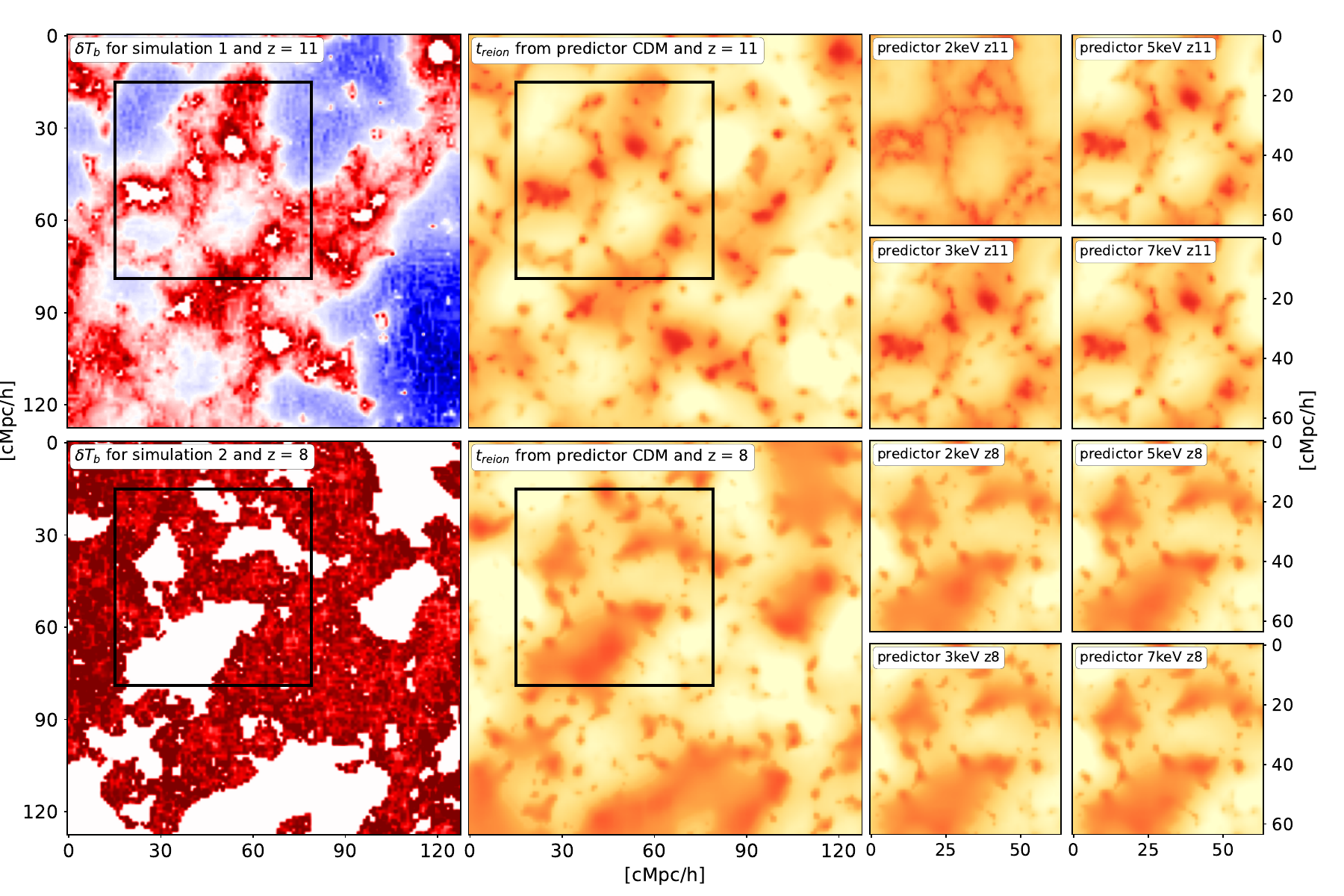}
    \caption{Example of predictions of $\TR$ for each predictor when the mock observations (input) is 21-cm maps from the model CDM. The left column is the 21-cm maps for $z=11$ (top) and $z=8$ (bottom). All the $\TR$ maps are predictions using a given predictor (written on the maps). \textcolor{black}{Only the CDM reconstructions are shown over the full field, whereas the reconstructions for the other predictors are shown for the region highlighted by the black square.}
}
    \label{fig:Predictions_maps}
\end{figure*}

\section{WDM predictors versus CDM inputs}
\label{sec:4_WDM}
Now, the idea is to investigate the behaviour of predictors when some data (that should describe the reality, the reference) are given to them. The ideal scenario is when the predictor model aligns with the model of the mock observation (see previous section). \textit{In this configuration, predictions of the reionisation time field should be statistically identical regardless of the redshift of observation $\zo$}. However, when the predictor model does not match the mock observation model, inconsistencies may emerge between reionisation time reconstructions at two distinct $\zo$, indicating a disparity between the predictor model and the "observation". A key approach to achieve this objective involves comparing two reconstructions at different $\zo$ for a given predictor model.

Then, from this point in the study, the CDM model is considered as the one to generate mock 21-cm observations: it is the reference and the only input for trained predictors, even when their underlying model does not match the reference model. The figure \ref{fig:pipelineWDM} shows the pipeline used to reconstruct $\TR$ and estimate the coherence of our CNN predictors (also see below \ref{sec:Model_Exclusion}). Thus, in Fig. \ref{fig:Predictions_maps}, examples of $\TR$ predictions for each predictor are presented where the input is the CDM 21-cm maps. In the top row, which represents predictions for $\zo=11$, it seems that the predictions for predictors associated with the models 3 keV, 5 keV, 7 keV, and CDM appear consistent. At this stage, solely relying on the maps, it becomes challenging to discern any notable differences between these reconstructions. However, the predictor linked to 2 keV model yields a $\TR$ map that substantially differs from the other predictors. This outcome was anticipated, given that 2 keV represents an exceptionally light particle mass, resulting in significant discrepancies in the 21-cm signal (input of the CNN) at this redshift. Moving to $\zo=8$ in the bottom row, all reconstructions exhibit similar features, including the 2 keV predictor. Consequently, no conclusions can be drawn from these maps at this point.

\subsection{Evaluating predictor self-consistency between $z = 11$ and $z = 8$}
\label{sec:Model_Exclusion}
One approach to distinguish and potentially exclude a predictor, thus its associated model, is to compare two reconstructions obtained from predictors at two different redshifts and the same DM model. In the subsequent section, we will use the same metrics ($Q_{\mathrm{HI}}$, P$_k$, isocontour length and minima number) employed in section \ref{sec:Models} to characterise $\TR$ predictions. For each metric, we will display the metric itself for each predictor, followed by the difference between both redshifts. This comparative analysis will provide insight into how the predictions evolve across different redshifts and may help in assessing the consistency and validity of each prediction model with respect to the observation model (CDM). In addition to the figures, several values are proposed to quantify the predictors' reconstruction consistency and are defined as follows.

Let any statistic $S_z$ computed on the reconstruction obtained with a predictor at redshift $z$ and evaluated over N points (e.g. $Q_{HI}^{z=8}$ or $P_k^{z=11}$). We define $X$ as the difference in this statistic between two redshifts given by $X=S_{11}-S_{8}$. The first value we want to estimate is the root mean square error (RMSE), which measures the magnitude of the average deviation relative to the zero line $X_0$. This zero line represents the expected value of $X$, as both redshifts should yield similar statistical properties within the reionisation time field. The RMSE is defined as:
\begin{equation} 
\label{eq:rmse}   
    RMSE=\sqrt{\langle (X-X_0)^2\rangle} = \sqrt{\langle X^2\rangle},
\end{equation}
where $X$ is a curve derived from the reconstructions and the operator $\langle \cdot \rangle$ represents the mean value. Hence, the RMSE represents the root-mean-square error of X between the reconstructions of the statistic $S$ at $z=11$ and $z=8$. Lower is the value of RMSE, better is the self-consistency for the statistic S. Additionally, for each RMSE value, we compute the coherence fraction q (see Tab. \ref{tab:tab}) which is simply defined as: 
\begin{equation}
q=\mathrm{RMSE_{CDM}}/\mathrm{RMSE_{WDM}}_i,
\end{equation}
normalising the WDM predictor error with the CDM predictor error. A q-value of 1 indicates identical RMSE values, while a q-value below 1 implies that the WDM predictor has a larger deviation on the statistic S when comparing reconstructions at $\zo=11$ and $\zo=8$. Conversely, a q-value greater than 1 would suggest that the WDM predictor has smaller deviation than the CDM one. Even though this latter outcome is not expected, as a predictor not trained to infer $\TR$ with CDM mock observations should not perform better than the CDM predictor, we may encounter situations where, for an incorrect predictor, the reconstructions for $\zo=11$ and $\zo=8$ are similar yet inconsistent with the ground truth, leading to a near-zero RMSE value and potentially a high q-value. In these circumstances, we cannot rely entirely on this single estimator. 
Finally, the $R^2_{\Delta}$ coefficient is also used as estimator and is defined as :
\begin{equation}
   R^2_{\Delta} = 1 - 
   \frac{\Sigma_{n}^{N} (X_{\mathrm{CDM}} - X_{\mathrm{WDM}})^2}
   {\Sigma_{n}^{N} (X_{\mathrm{CDM}} - <X_{\mathrm{CDM}}>)^2},
\end{equation}
where $X_{\mathrm{WDM}}$ and $X_{\mathrm{CDM}}$ are the reconstructions' comparison with respect to a given WDM predictor and the CDM predictor, respectively. The quantity $<X_{\mathrm{CDM}}>$ displays the mean value of $X_{\mathrm{CDM}}$ computed over the N evaluation points of $X$, and by extension, of $S$. 
The $R^2_{\Delta}$ value quantifies the distance between each CDM and WDM reconstruction consistency ($X_{\mathrm{CDM}} - X_{\mathrm{WDM}}$) relative to the distance of each CDM reconstruction to its mean ($X_{\mathrm{CDM}}-\langle X_{\mathrm{CDM}} \rangle$), and we have:
\begin{itemize}
    \item if $R^2_{\Delta}<0$: The WDM reconstruction is more distant from the corresponding CDM reconstruction than the latter is from the mean CDM reconstruction. Thus, WDM is less consistent than CDM, which suggests that the underlying model is not suited to the input data. 
    \item if $R^2_{\Delta}=0$: The WDM reconstruction is as distant as the mean CDM reconstruction from the corresponding CDM reconstruction.
    \item if $0<R^2_{\Delta}\leq1$: The WDM reconstruction is closer to the corresponding CDM reconstruction than the latter is to its mean value. The larger the $R^2_{\Delta}$ value is, the more the WDM reconstruction is consistent with the CDM reconstruction. The extreme case being $R^2_{\Delta}=1$, where each WDM reconstruction has the same consistency as the corresponding CDM reconstruction.    
\end{itemize}
\begin{figure}
    \centering
    \includegraphics[width=0.5\textwidth]{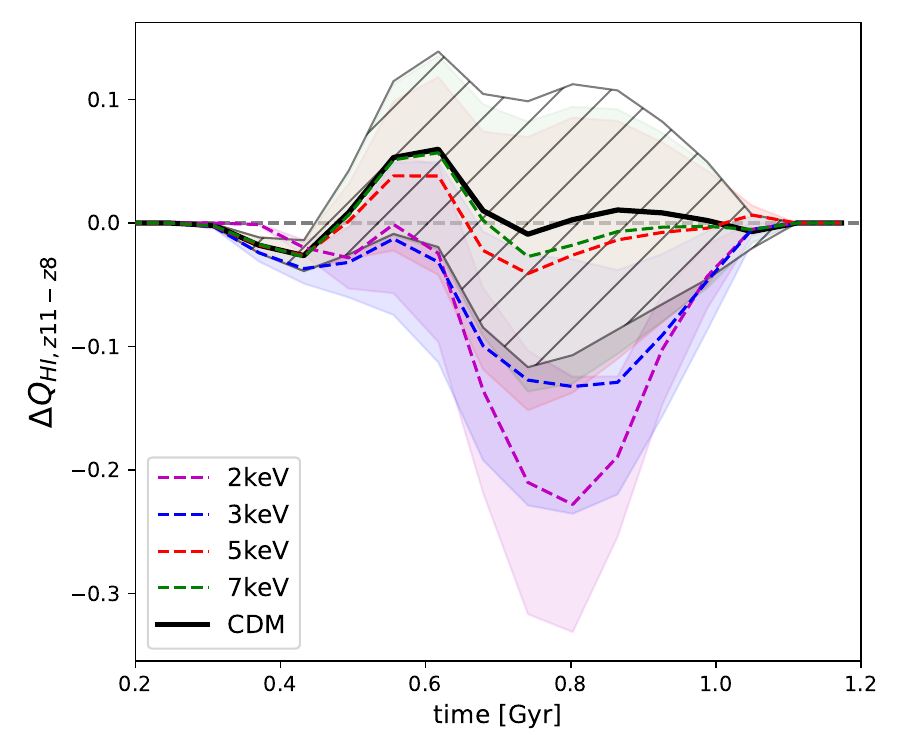}
    \caption{Comparison of the neutral hydrogen volume fraction between redshift 11 and 8 for each model. Dashed/shaded areas stand for the standard deviation, while solid lines depict the mean.}
    \label{fig:QHI11-QHI8}
\end{figure}
\subsection{Neutral fraction - $Q_{\mathrm{HI}}$}
\label{subsec:QHI}
The volume fraction of neutral hydrogen $Q_{\mathrm{HI}}$ is the first metric we want to investigate. The raw results are shown in appendix \ref{app:B:QHI} and the figure \ref{fig:QHI11-QHI8} illustrates $\Delta Q_{\mathrm{HI}}$, the difference between the predicted volume neutral fraction at $z=11$ minus $z=8$ for each model. Ideally, this difference should be zero. However, the CDM predictor, which aligns with the mock observation model, shows deviations ranging between -2$\%$ and 6$\%$ from perfect correlation.  The 5 and 7 keV predictors demonstrate deviations within the range of [-4$\%$, 6$\%$]. In contrast, the 2 and 3 keV predictors show larger deviations, with absolute values of up to 13$\%$ and 23$\%$, respectively. This discrepancy suggests that these two predictors are not consistent with the model of the mock observation, and provides an initial hint that they can be potentially excluded from acceptable models. Furthermore, from Tab. \ref{tab:tab}, the RMSE and $R^2_{\Delta}$ values confirm this trend. For masses 5 and 7 keV, the $R^2_{\Delta}$ value is positive, 0.38 and 0.83 respectively, suggesting that the three curves (including CDM) follow the same behaviour. However, the $R^2_{\Delta}$ value is negative for the 2 and 3 keV masses, hinting once again that these two predictors do not align with the others.
\textcolor{black}{The corresponding epistemic uncertainty ratios $\Delta\sigma$ (cf. Table~\ref{tab:sigma}) are consistently found to be below unity for all predictors, with values ranging from 0.24 to 0.29 for the max value. Further interpretation of these results is provided in Sec.~\ref{sec:Discussion}.}
\subsection{Reionisation time power spectrum}
\label{subsec:Pk}
The raw power spectra for both redshifts are presented in the appendix \ref{app:B:Pk}.
The figure \ref{fig:Pk11_vs_Pk8} displays $\Delta P_k$ the relative $P_k$ differences between $z=11$ and $z=8$, normalised by dividing it by $P_k$ at $z=8$. The self-consistencies of the 5, 7 keV and CDM predictors follow the same trend and are pretty close to each other: they are all within the range [-16, 13]$\%$ deviation from the perfect correlation. The 2 and 3 keV predictors do not follow the same behaviour and deviate more: [-51, -16]$\%$ and  [-24, 10]$\%$ for 2 and 3 keV respectively. At this point, taking aside the 2 keV predictor, it becomes tricky to know whether a predictor can be excluded or not: the 3 keV predictor seems to present a higher level of inconsistency than the other predictors, but it is not sufficient to accurately exclude it. In Tab. \ref{tab:tab} the RMSE, q and $R^2_{\Delta}$ values for $\Delta P_k$ are presented. As previously for the neutral fraction, the q-values for the 5 and 7 keV predictors, 1.02 and 0.97 respectively, suggest a reconstruction similar to the CDM. Additionally, the $R^2_{\Delta}$ values for these predictors (0.93 and 0.94) indicate a strong correlation between their reconstructions and those of the CDM. These values demonstrate the difficulty in distinguishing these models based on the power spectrum alone. On the other hand, the q and $R^2_{\Delta}$ values for the 3 keV model are 0.61 and -0.068 in this situation. The negative value of $R^2_{\Delta}$ suggests an incompatibility with the CDM input.
\textcolor{black}{The corresponding epistemic uncertainty ratios $\Delta\sigma$ for the power spectrum remain below unity for all predictors. The values are even lower than those obtained for $Q_{\mathrm{HI}}$ ($\Delta\sigma_{max} \approx 0.1$), indicating that the reconstruction uncertainty associated with the CNN is subdominant with respect to the intrinsic scatter of the reconstructed power spectrum.}
\begin{figure}
    \centering
    \includegraphics[width=0.5\textwidth]{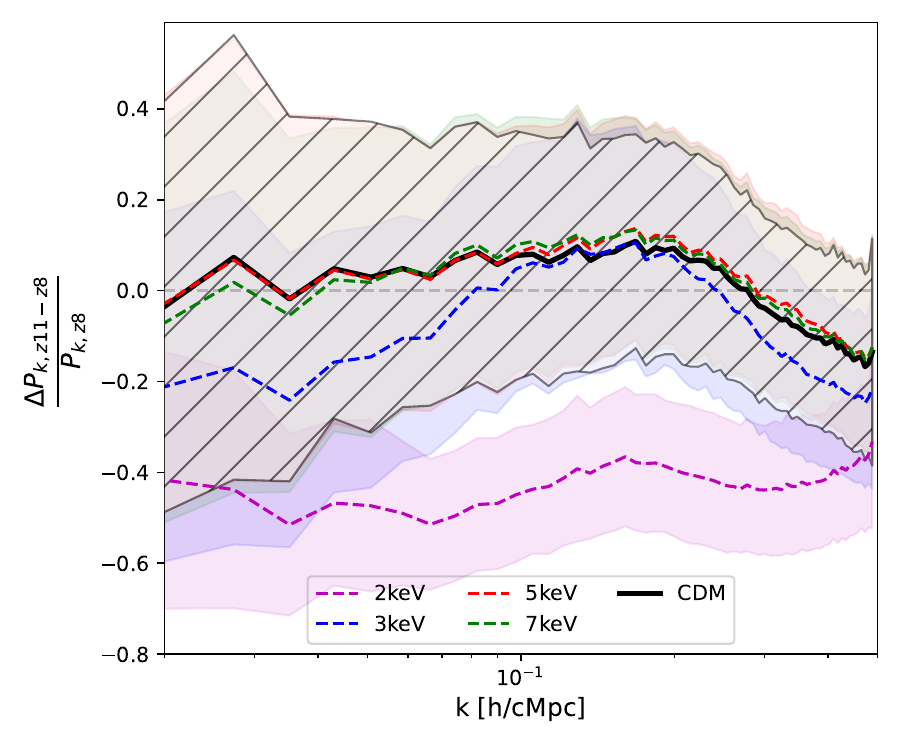}
    \caption{Comparison of the $\TR$ power spectrum between redshift 11 and 8 for each model. Dashed/shaded areas stand for the standard deviation, while solid lines depict the mean.}
    \label{fig:Pk11_vs_Pk8}
\end{figure}
\subsection{Isocontour Length}
\label{subsec:L}
\begin{figure}
    \centering
    \includegraphics[width=0.5\textwidth]{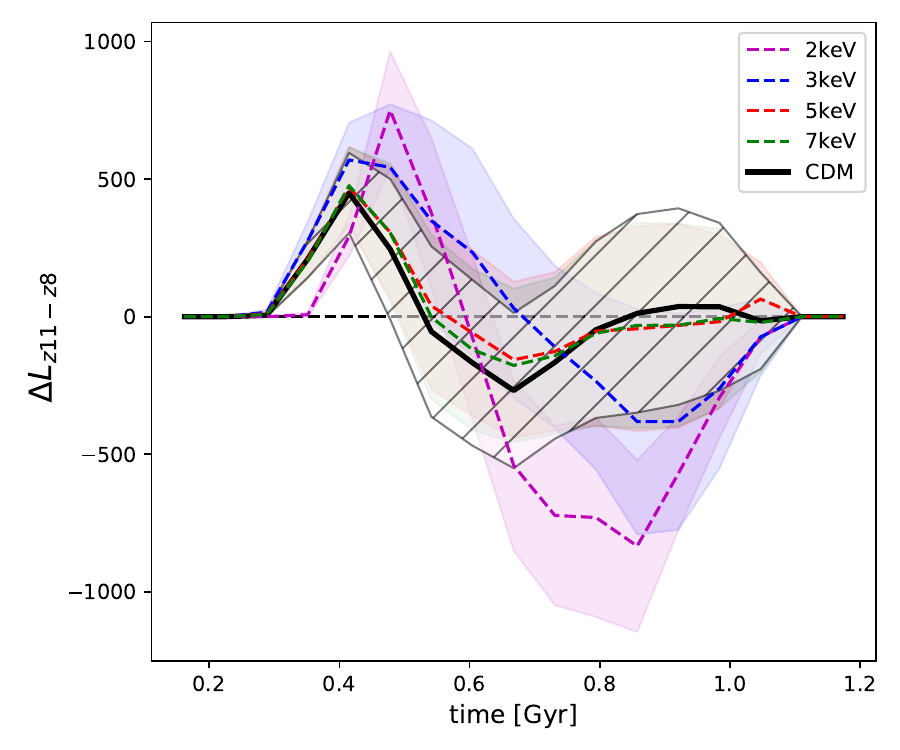}
    \caption{Comparison of the total isocontour length between redshift 11 and 8 for each model. Dashed/shaded areas stand for the standard deviation, while solid lines depict the mean.}
    \label{fig:L11_vs_L8}
\end{figure}

The next metric is the total isocontour length, as previously introduced in Section \ref{sec:Models}, and the raw results are presented in the appendix \ref{app:B:L}.
The differences for each predictor between redshifts 11 and 8 are shown in Fig. \ref{fig:L11_vs_L8}. 
For the 5, 7 keV and CDM models, the predictor self-consistency follows again a similar trend: they predict more isocontour length from mock observation at $z=11$ than at $z=8$ before $t\approx0.55$ Gyr and then fewer lengths at later times. The 5 and 7 keV models closely align within $1 \sigma$ (shaded area) of the CDM, meaning the impossibility of properly distinguishing these models at this stage of the study. 
Looking at the Tab. \ref{tab:tab}, the RMSE for the 2 and 3 keV predictors is larger compared to the 5, 7 keV and CDM predictors, 
In consequence, the q-value is similar for the masses 5 and 7 keV ($\approx$ 1) while it drops to 0.58 for 3 keV. The $R^2_{\Delta}$ coefficients exhibit similar values to the ones in the previous section leading to a comparable conclusion: The 2 and 3 keV predictors ($R^2_{\Delta} = -4.8$ and $-1.2$, respectively) have a larger level of inconsistency compared to the other models ($R^2_{\Delta} = 0.87$ and $0.94$, for the models 5 keV and 7 keV, respectively) because their behaviours deviate from the other predictors, showing that these two predictors lost similarities with the reference and can be excluded from isocontours stats. 
\textcolor{black}{The values of $\Delta\sigma$ remain below unity across all predictors. Although the absolute values vary slightly between models, ranging from 0.13 to 0.20 for $\Delta\sigma_{\mathrm{max}}$, the epistemic uncertainty of the reconstruction remains significantly smaller than the dispersion of $L$ across the test set.}
\subsection{Minima - Reionisation sources}
\label{subsec:N}
\begin{figure}
    \centering
    \includegraphics[width=0.5\textwidth]{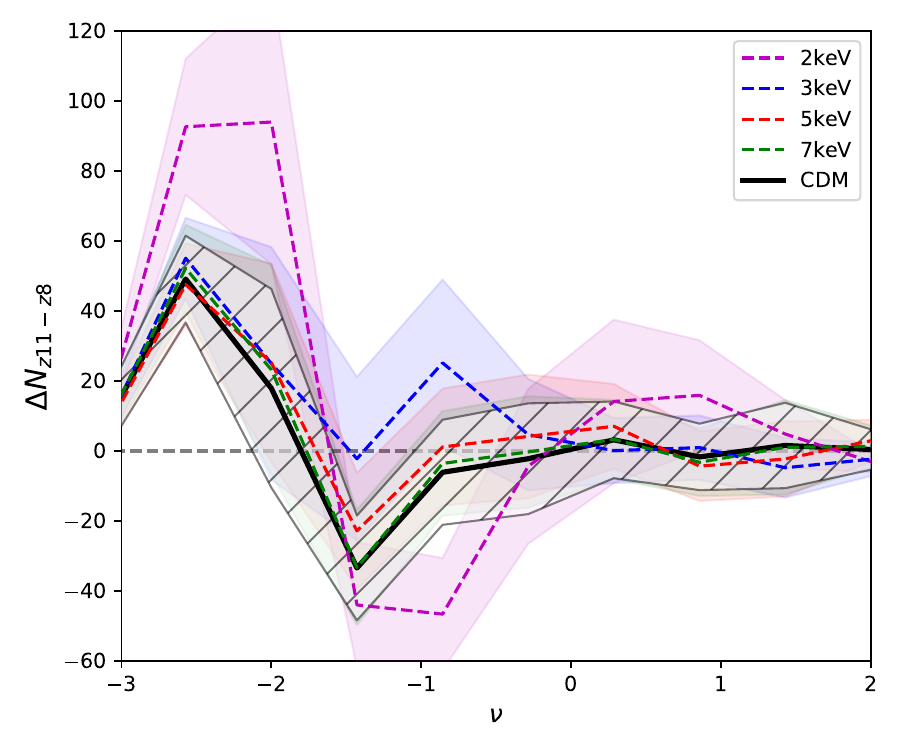}
    \caption{Comparison of the minima number between redshift 11 and 8 for each model. Dashed/shaded areas stand for the standard deviation, while solid lines depict the mean.}
    \label{fig:N11_vs_N8}
\end{figure}

The last quantity we want to investigate is the number of seeds N, which indicates the number of minima of the $\TR$ field. This quantity holds significance as it shows the first regions to undergo reionisation, essentially the first reionisation sources. Accurately locating these regions is therefore crucial for studying the oldest galaxies in our universe, allowing us to analyse their formation and role in reionising their surroundings. Nevertheless, it is worth noting that the images used in this paper have a $1\mpc$/pixel resolution, which is not sufficient to capture such small-scale features but can still allow us to locate the sources. As for the other metrics, the raw results are presented in the appendix \ref{app:B:N}.
The difference $N_{z11} - N_{z8}$ is shown in Fig. \ref{fig:N11_vs_N8}. Once again, the behaviour of the 5, 7 keV and CDM curves is similar, with the former two located within the latter's standard deviation and having a comparable q-value (Tab. \ref{tab:tab}): 1.05 and 0.94, for the 5 and 7 keV predictors, respectively. Surprisingly, the 3 keV predictor reaches a $q$-value of 0.97, indicating that its behaviour exhibits a similar level of inconsistency as the CDM predictor. Conversely, the 2 keV predictor can be ruled out based on this single statistic: it predicts a significantly large number of early sources at z=11 due to the toroidal structures found in the prediction. As a consequence, a large peak is observed in the Fig. \ref{fig:N11_vs_N8} at $\nu_{\TR}\approx$-2 and its coherence fraction $q = 0.41$. In terms of $R^2_{\Delta}$, only the 2 keV predictor displays a negative value. Now, the 3 keV predictor shows a positive $R^2_{\Delta}$ value for this metric, yet small (0.53) against 0.94 and 0.98 for the 5 and 7 keV predictors, respectively. Despite its high $q$-value, the 3 keV predictor does not exhibit the same behaviour as the CDM predictor (small $R^2_{\Delta}$), once again suggesting that this model is less well suited to reproduce the input model.
\textcolor{black}{For the statistic $N$, $\Delta\sigma$ remains also below unity for all predictors. This confirms that the uncertainty introduced by the CNN reconstruction remains low compared to the intrinsic variability of the observable, despite slightly larger values for this statistic compared to the others.}

\section{Discussion}
\label{sec:Discussion}
In this study, we proposed a direct comparison of each predictor's performance against the expected behaviour of the CDM predictor, by feeding all predictors with CDM 21-cm maps. However, since each predictor is trained on a specific dark matter model, it inherently reflects the expected reconstructions of that model. 
In addition, in practice, the underlying dark matter model of an observation is unknown, meaning that the reconstruction of any predictor cannot be directly compared to an input reference, since the true model is not known. A more rigorous approach would therefore consist of evaluating each predictor within its own theoretical framework. For instance, when applying a WDM-trained predictor to CDM input maps, its reconstruction should be compared not to the CDM reference, but to the expected reconstructions predicted by the WDM model itself when fed by its own input model. This would allow us to quantify how much the predictor deviates from its intrinsic expectations rather than from another model’s behaviour. 
Nevertheless, the performance of the predictors (with aligned input) appears consistent in our analysis, which means that the predicted field $\TR$ exhibits similar properties across different models. This consistency suggests that using CDM as a reference does not substantially impact the ability to distinguish between models at first order. The closer a model is to the reference, the harder it becomes to differentiate it when comparing it to the reference. \textcolor{black}{A complementary perspective is presented in Appendix \ref{app:C:WDMref}, where the 2 keV model is adopted as the mock observation.}

Furthermore, an important subtlety emerges when examining the results for the 5 KeV predictor: the RMSE value for this predictor is slightly below the RMSE for the CDM one, resulting in a q-value above 1 for all the statistics presented. As said previously, this behaviour is not expected since the 5 keV predictor is not trained on CDM inputs. Although the differences in terms of RMSE remain quite small and are not significant in this case, we may encounter situations where a "wrong" predictor yields a better RMSE value than a "right" predictor, raising critical questions. For example, a predictor that reconstructs an arbitrary statistic $S$ similarly at two different redshifts may produce a small $X$ (small RMSE value), suggesting an apparently coherent reconstruction. However, this outcome can be misleading: for instance, two identical predictions at different redshifts, yet deviating significantly from the ground truth, would still result in a $X$ close to zero. Thus, the magnitude of $X$ alone is insufficient to assess the quality of the predictor, reinforcing the necessity of direct comparisons with a trusted reference, here the CDM expectation: This is the reason we proposed the $R^2_{\Delta}$ estimator to compare the reconstructions to the expected behaviour of CDM. \textcolor{black}{Moreover, the epistemic uncertainty analysis based on Monte Carlo dropout (Sect.~\ref{sec:4_WDM}) shows that the uncertainty induced by the CNN reconstruction remains systematically subdominant, with $\Delta\sigma < 1$ for all observables and predictors. This indicates that the model-induced noise does not significantly contaminate the statistical comparisons presented in this work. However, the exact magnitude of $\Delta\sigma$ varies across observables, suggesting that our CNN does not reconstruct all metrics with the same level of precision.}

Also, the primary objective of this study is not to constrain cosmological/astrophysical parameters, but to assess the robustness and behaviour of predictors when faced with observational data inconsistent with their training model. Our results show that it is indeed possible to distinguish whether a predictor's model aligns with the observational input. A model is thus considered valid not because it is necessarily true, but because it contains a set of parameters capable of reproducing the observational features. Ultimately, a combination of different parameters, and thus models, may satisfy observational constraints, and further cross-correlation with independent studies will be necessary to progressively rule out or refine viable parameter spaces. \textcolor{black}{In addition, as stated in Sect. \ref{sec:Models}, the ionising efficiency plays a role similar to that of the WDM particle mass in shaping the reionisation history (along with other parameters not explored in this study), leading to a highly degenerate parameter space. Therefore, our results do not imply that the CNN is specifically sensitive to the WDM model, but rather to variations in the 21-cm signal and the resulting reionisation history.}

Eventually, the ability to discern models based on predictors is, however, highly dependent on the CNN's performance. Improving the CNN architecture is a natural next step to enhance model selection capabilities. 
Additionally, refining the resolution of training simulations could further boost predictive accuracy. Nevertheless, our ultimate goal is to develop tools suited to the analysis of SKA-like observations, which will inevitably be characterised by limited resolution and contaminated by instrumental noise and foregrounds.

\begin{table*}
\centering
\setlength{\tabcolsep}{10pt} 
\begin{tabular}{c|cc|cc|cc|cc}
\toprule
\multirow{2}{*}{\textbf{Predictor}} &
  \multicolumn{2}{c|}{\textbf{Q$_{\textbf{HI}}$}} &
  \multicolumn{2}{c|}{\textbf{P$_\textbf{k}$}} &
  \multicolumn{2}{c|}{\textbf{L}} &
  \multicolumn{2}{c}{\textbf{N}} \\
& \multicolumn{1}{c}{\textbf{RMSE / q}} & \multicolumn{1}{c|}{\textbf{$R_{\Delta}^2$}} &
  \multicolumn{1}{c}{\textbf{RMSE / q}} & \multicolumn{1}{c|}{\textbf{$R_{\Delta}^2$}} &
  \multicolumn{1}{c}{\textbf{RMSE / q}} & \multicolumn{1}{c|}{\textbf{$R_{\Delta}^2$}} &
  \multicolumn{1}{c}{\textbf{RMSE / q}} & \multicolumn{1}{c}{\textbf{$R_{\Delta}^2$}} \\
\midrule
\textbf{2 keV} & 0.090 / 0.22 &-2.3e+01& 0.422 / 0.21 & -2.2e+01& 4.4e+02 / 0.37 & -4.8 & 36.5 / 0.45 & -0.95 \\
\textbf{3 keV} & 0.061 / 0.32 &-1.1e+01& 0.141 / 0.61 & -0.068  & 2.8e+02 / 0.58 & -1.2 & 17.1 / 0.97 & 0.53 \\
\textbf{5 keV} & 0.019 / 1.06 & 0.38   & 0.087 / 1.02 & 0.93    & 1.6e+02 / 1.03 & 0.87 & 15.7 / 1.05 & 0.92 \\
\textbf{7 keV} & 0.020 / 0.99 & 0.83   & 0.090 / 0.97 & 0.94    & 1.6e+02 / 1.01 & 0.94 & 17.6 / 0.94 & 0.99 \\
\textbf{CDM}  & 0.020 / 1.0  & 1.0    & 0.088 / 1.0  & 1.0     & 1.6e+02 / 1.0  & 1.0   & 16.6 / 1.0  & 1.0 \\
\bottomrule
\end{tabular}
\caption{Inconsistency estimators of the predictors. RMSE, q and $R_{\Delta}^2$ are the root-mean-square error, the coherence fraction and the determination coefficient as detailed in the Sec. \ref{sec:Model_Exclusion}. For a consistent predictor, the RMSE should be zero. The values of q and $R_{\Delta}^2$ are computed with respect to the CDM, thus for a predictor with the same behaviour (same consistency level) as the right predictor (CDM), these values should be 1.}
\label{tab:tab}
\end{table*}

\begin{table*}
\centering
\setlength{\tabcolsep}{10pt} 
\begin{tabular}{c|cc|cc|cc|cc}
\toprule
\multirow{2}{*}{\textbf{Predictor}} &
  \multicolumn{2}{c|}{\textbf{Q$_{\textbf{HI}}$}} &
  \multicolumn{2}{c|}{\textbf{P$_\textbf{k}$}} &
  \multicolumn{2}{c|}{\textbf{L}} &
  \multicolumn{2}{c}{\textbf{N}} \\
& \multicolumn{1}{c}{\textbf{$\langle\Delta\sigma\rangle$}} & \multicolumn{1}{c|}{\textbf{$\Delta\sigma_{\mathrm{max}}$}} &
  \multicolumn{1}{c}{\textbf{$\langle\Delta\sigma\rangle$}} & \multicolumn{1}{c|}{\textbf{$\Delta\sigma_{\mathrm{max}}$}} &
  \multicolumn{1}{c}{\textbf{$\langle\Delta\sigma\rangle$}} & \multicolumn{1}{c|}{\textbf{$\Delta\sigma_{\mathrm{max}}$}} &
  \multicolumn{1}{c}{\textbf{$\langle\Delta\sigma\rangle$}} & \multicolumn{1}{c}{\textbf{$\Delta\sigma_{\mathrm{max}}$}} \\
\midrule
\textbf{2 keV} & 0.09 & 0.27 & 0.08 & 0.11 & 0.1 & 0.20 & 0.24 & 0.42 \\
\textbf{3 keV} & 0.07 & 0.23 & 0.07 & 0.09 & 0.8 & 0.13 & 0.29 & 0.38 \\
\textbf{5 keV} & 0.09 & 0.29 & 0.07 & 0.11 & 0.9 & 0.17 & 0.26 & 0.42 \\
\textbf{7 keV} & 0.08 & 0.27 & 0.06 & 0.10 & 0.8 & 0.15 & 0.24 & 0.31 \\
\textbf{CDM}   & 0.09 & 0.24 & 0.07 & 0.10 & 0.9 & 0.18 & 0.24 & 0.33 \\
\bottomrule
\end{tabular}
\caption{\textcolor{black}{Epistemic uncertainty ratios $\Delta\sigma = \sigma_X$/$\mathrm{std}_X$ for the different predictors and observables (see Sec.~\ref{sec:TR_reconstruction}). For each quantity, we report the mean value $\langle \Delta\sigma \rangle$ and the maximum value $\Delta\sigma_{\mathrm{max}}$ over the test dataset. The values are shown for the four observables considered in this work: the neutral fraction $Q_{\mathrm{HI}}$, the power spectrum $P_k$, the total isocontour length $L$ and the minima $N$ (see main text for definitions).}}
\label{tab:sigma}
\end{table*}

\section{Conclusion}
\label{sec:conclusion}
In this paper, we proposed a method to evaluate the coherence of our CNN predictors with respect to their input model, focusing on the reconstruction of the reionisation time field $\TR$. 
Building on previous work demonstrating the feasibility of predicting $\TR$ from 21-cm observations at a given redshift, we specifically address a critical limitation: the fact that a predictor is trained on a given reionisation model, which may not perfectly match reality. We propose to measure the coherence of the predictor with CDM inputs, by evaluating metrics such as the (RMSE), the coherence fraction $q$, and the coefficient of determination $R_{\Delta}^2$ between predictions at different redshifts.

Applying this methodology to WDM models with particle masses of 2, 3, 5, and 7 keV, we find that the 5 and 7 keV predictors show a level of coherence comparable to that of the CDM predictor when applied to the CDM input model. These higher mass models have similarities with the CDM model, making it challenging to distinguish them. In contrast, the 2 keV predictor coherence exhibits significant deviation, particularly in the $Q_{HI}$ and $P_k$ statistics. The 2 keV model has already been ruled out in previous studies (e.g. \cite{Irsic_2024}) due to its substantial departure from the CDM model, allowing it to be more easily excluded. Finally, the 3 keV predictor, while also ruled out, displays an intermediate behaviour, which complicates the assessment of his underlying model's validity when fed with CDM input. Our analysis highlights that even predictors showing consistent behaviour on individual metrics (e.g. the 3 keV predictor) must be carefully evaluated across multiple quantities to validate or invalidate underlying models. 

Importantly, we demonstrated that model discernibility is directly tied to the CNN's performance. Although current results are promising, further improvements are necessary, especially in anticipation of SKA observations which will be characterised by lower resolution and observational noise. Future work will therefore focus on enhancing CNN architectures, for example, by implementing attention mechanisms or transformer-based designs, and developing methods for denoising and interpreting real observational data.

By providing a framework for testing predictor coherence in $\TR$ reconstructions, this work paves the way for more reliable applications of machine learning to the Epoch of Reionisation, ultimately contributing to a better understanding of cosmic reionisation and the nature of dark matter.

\section*{Acknowledgement}
The authors would like to acknowledge the High-Performance Computing Centre of the University of Strasbourg for supporting this work by providing scientific support and access to computing resources. Part of the computing resources were funded by the Equipex Equip@Meso project (Programme Investissements d’Avenir) and the CPER Alsacalcul/Big Data. This work was granted access to the HPC resources of TGCC under the allocations 2023-A0130411049 “Simulation des signaux et processus de l’aube cosmique et Réionisation de l’Univers” made by GENCI.
The authors acknowledge funding from the European Research Council (ERC) under the European Unions Horizon 2020 research and innovation programme (grant agreement No. 834148).

%
\bibliographystyle{aa} 
\bibliography{biblio.bib} 
%


\begin{appendix}
\section{Prediction with aligned models}
\label{app:A}
\begin{figure*}
    \centering
    \includegraphics[width=1\textwidth]{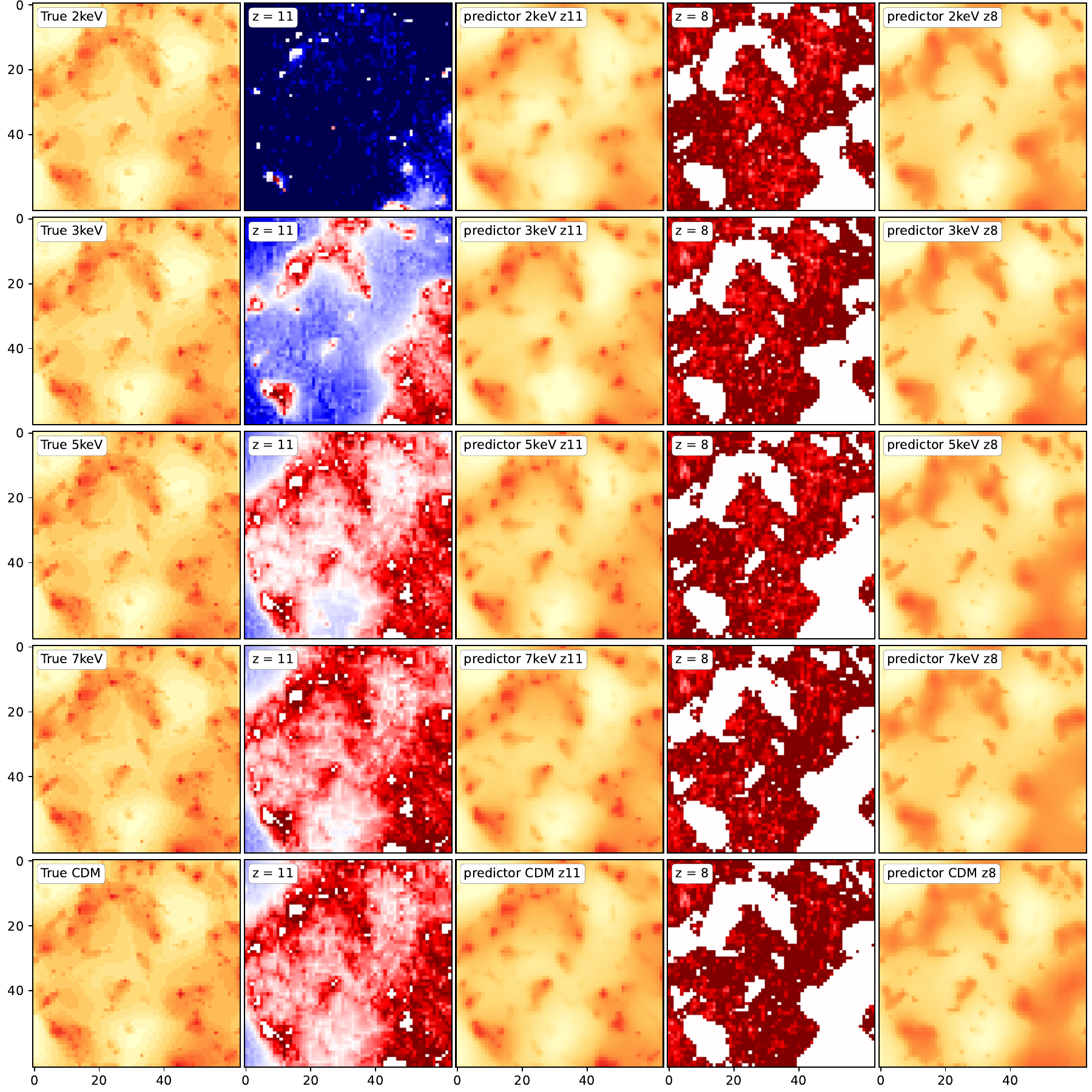}
    \caption{\textcolor{black}{$64\times64$ $\mpc$ subregion of $\TR$ Predictions for the five DM models when the mock observation (input) matches with the predictor model.} Each row represents a given model, from top to bottom: 2 keV, 3 keV, 5 keV, 7 keV, and CDM. The first column depicts the ground truth that predictors aim to infer. $2^{nd}$ and $4^{th}$ columns are for the 21-cm maps at $z=11, 8$ respectively. On their right ($3{rd}$ and $5{th}$ columns) their respective predictions are shown. The unit of x and y-axis is [$\mpc$].}
    \label{fig:pred_right_obs}
\end{figure*}

Before testing predictors on mismatched models, we need to make sure that each predictor performs similarly when they are submitted to their own observation model: a predictor should not over-perform or under-perform the others. 

Figure \ref{fig:pred_right_obs} illustrates the $\TR$ predictions when each predictor is fed by 21-cm maps that align with its respective underlying model. Each row corresponds to a distinct model, which features the ground truth $\TR$, the 21-cm maps at both redshifts ($\zo$ = 11 and 8) and the corresponding predictions $\TR$. It is noteworthy that here the two 21-cm maps (from $z=11$ and $z=8$) depict the same region at different points in time, but this will not be the case in future figures.
The third column exhibits the $\TR$ predictions for each predictor using 21-cm maps at $\zo=11$ (second column). Regardless of the predictor model, the overall reconstruction quality remains consistent: large-scale structures are accurately represented, and smaller scales, such as approximately 5 $\mpc$, are discernible. Nonetheless, the predictions exhibit a smoother resolution compared to the ground truth.
In the fifth column, predictions from $\zo=8$ input maps are depicted. Once again, the reconstruction quality remains fairly consistent between the predictors. At this redshift, the predictions appear slightly more blurred compared to those at redshift 11, but the overall representation of the reionisation history is satisfactory. 

The $R_2$ coefficients were computed and are shown in Tab. \ref{tab:R2}. The values are consistent, ranging from 0.85 to 0.9 for the five predictors at both redshifts. This suggests a comparable reconstruction performance, where no predictor outperforms the others significantly on this statistic. This primary analysis is crucial for avoiding significant bias, ensuring that no single predictor is disproportionately well-trained while another fails to properly recover $\TR$. Furthermore, it suggests that regardless of the dark matter model and deviations within the 21-cm signal (especially at $\zo$=11), inferring the reionisation time can be achieved similarly with an appropriate predictor. For further insights, refer to \cite{Hiegel_2023}, which delves into the performance evaluation of our CNN for CDM models.

\begin{table}[h]
    \centering
    \begin{tabular}{c||c|c|c|c|c}
        \toprule
        \diagbox{\textbf{\Large{z$_{\textbf{obs}}$}}}{\large{\textbf{model}}} & \textbf{2 keV} & \textbf{3 keV} & \textbf{5 keV} & \textbf{7 keV} & \textbf{CDM} \\
        \midrule
        \textbf{8} & 0.89 & 0.88 & 0.88 & 0.90 & 0.87\\
        \midrule
        \textbf{11} & 0.85 & 0.88 & 0.88 & 0.88 & 0.9 \\
        \bottomrule
    \end{tabular}
    \caption{Determination coefficient $R^2$ for the five predictors, and both redshift when the observation model aligns with the predictor. These values were computed using the whole test set (500 images).}
    \label{tab:R2}
\end{table}

\section{WDM predictors versus CDM inputs - raw results}
This appendix presents and analyses the raw results for each statistic (neutral fraction, power spectrum, isocontour length, and minima) for both redshifts.

\subsection{Neutral fraction - $Q_{\mathrm{HI}}$}
\label{app:B:QHI}
\begin{figure*}
    \centering
    \includegraphics[width=0.73\textwidth]{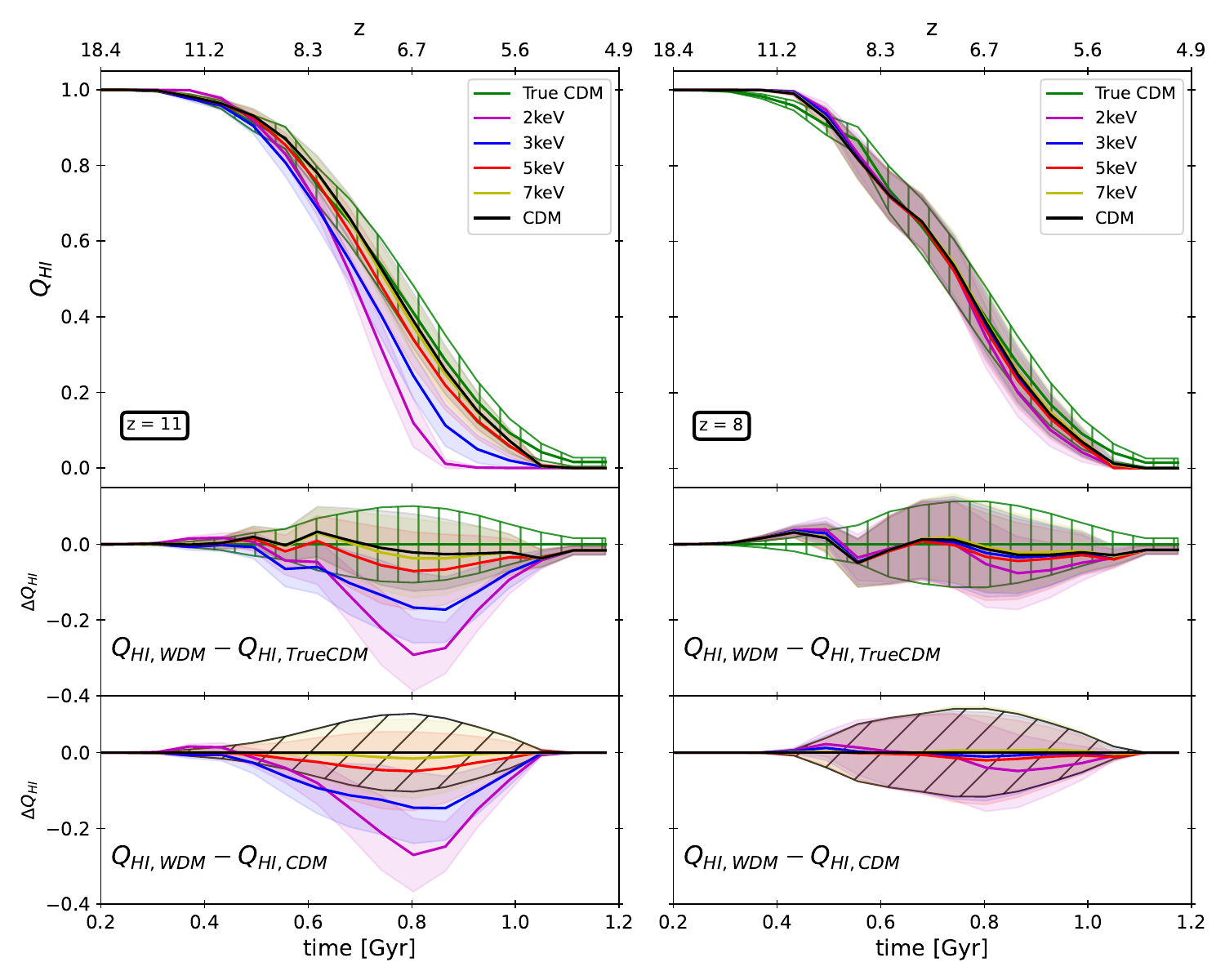}
    \caption{Volume fraction of neutral Hydrogen $Q_{\mathrm{HI}}$ for each predictor. Left and right are for $z=11$ and $8$ respectively. The top panel is the mean and standard deviation of $Q_{\mathrm{HI}}$ with, in green, the ground truth coming from the CDM model. The middle panel represents the difference between WDM models and the ground truth, and the bottom panel represents the difference between WDM models and the CDM prediction.}
    \label{fig:QHI_vs_CDM}
\end{figure*}
The volume fraction of neutral hydrogen $Q_{\mathrm{HI}}$ is depicted in Figure \ref{fig:QHI_vs_CDM}, the predicted curves for $z = 8$ on the right reveal no relevant differences between predictors. This outcome aligns with our expectations based on the maps previously shown (Fig. \ref{fig:Predictions_maps}), where the maps look similar, and the 21-cm maps at $z=8$ (Fig. \ref{fig:treion_tb_maps}), which exhibit similarities between models. However, on the left side, the predictions for z = 11 highlight some distinctions between predictors. Specifically, the 2 and 3 keV predictors fail to reconstruct a consistent reionisation history compared to the ground truth (depicted in green). At $t = 0.8$ Gyr, there is a 29$\%$ and 16$\%$ deficit for the 2 and 3 keV predictors, respectively, in comparison to the ground truth. Similarly, when compared to the CDM predictor (which serves as the reference), these predictors exhibit deficits of 27$\%$ and 14$\%$ at the same considered time.

\subsection{Reionisation time power spectrum}
\label{app:B:Pk}
\begin{figure*}
    \centering
    \includegraphics[width=0.73\textwidth]{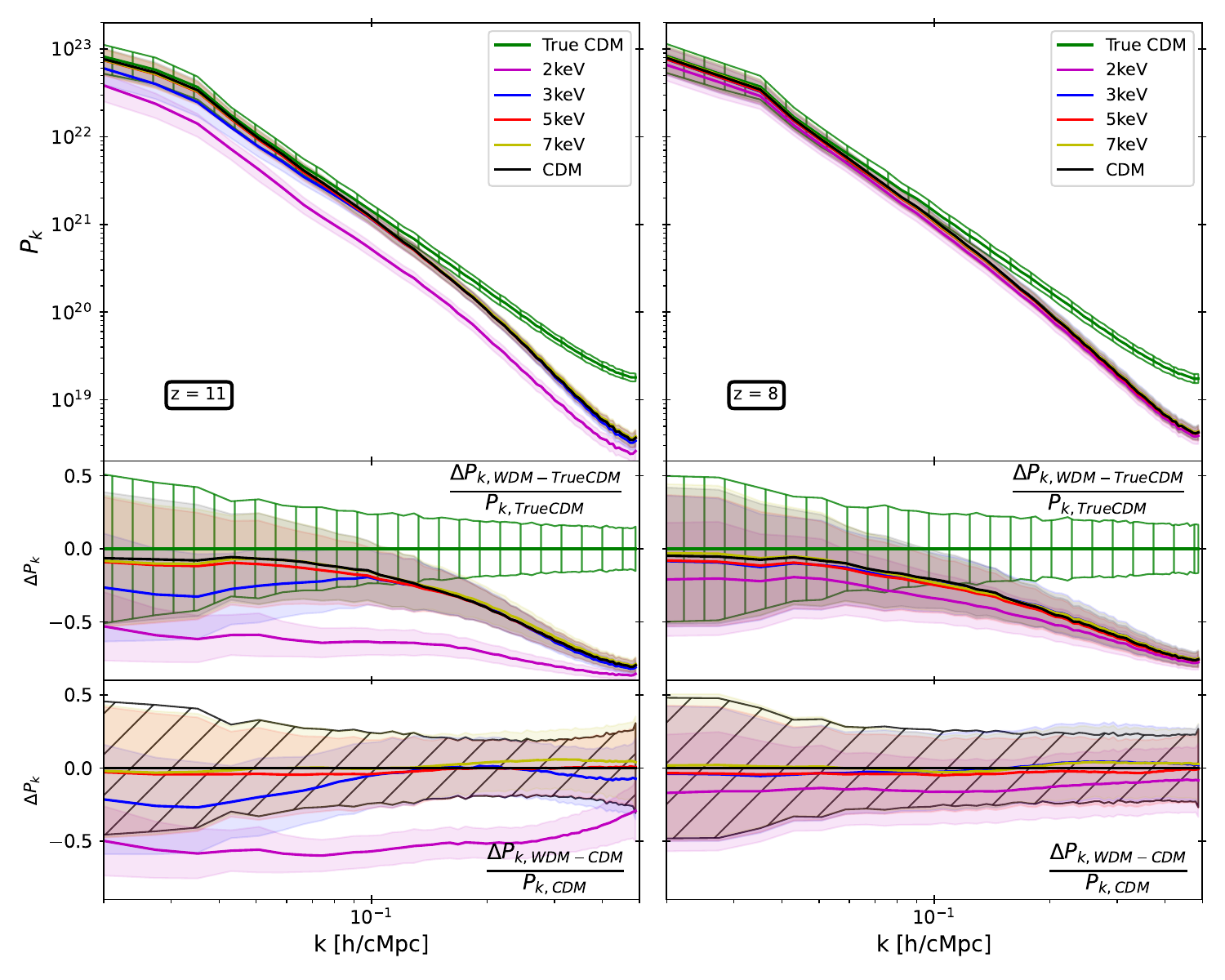}
    \caption{Power spectrum $P_{k}$ for each predictor. Left and right are for $z=11$ and $8$ respectively. The top panel is the mean and standard deviation of $P_{k}$ with, in green, the ground truth coming from the CDM model. The middle panel represents the difference between WDM models and the ground truth, and the bottom panel represents the difference between WDM models and the CDM prediction.}
    \label{fig:Pk_vs_CDM}
\end{figure*}
The predicted $\TR$ power spectrum $P_k$ is shown in Fig. \ref{fig:Pk_vs_CDM}. The CDM $P_k$ prediction looks identical regardless of the observation redshift: The large-scale structures ($k < 0.1$ $\hmpc$) are well represented and there is a drop of power in the prediction that completely goes off the error bars for the small-scale structures ($k > 0.1$ $\hmpc$). This phenomenon was already observed in \cite{Hiegel_2023}, and was expected looking at the predicted maps that look smoother than the ground truth. The predictor for 5 and 7 keV follows the same trend and are pretty close to the CDM: both are within $5\%$ deviation from the predicted CDM for both redshifts and the whole k range. The 3 keV predictor is also within the same deviation for $\zo=8$. For $\zo=11$, the deviation goes up to $9\%$ for $k > 0.3$ $\hmpc$ and up to 27$\%$ for $k < 0.1$ $\hmpc$: When dealing with CDM mock observation, the 3 keV predictor have difficulties to infer large scales structures (small spatial frequencies $k > 0.1$ $\hmpc$) but also smooths further the small scales ($k > 0.3$ $\hmpc$). Looking for the extreme scenario, the 2 keV predictor $P_k$ reconstruction has a huge deficit in power for the whole k range and both redshift: between $30$ and $60\%$ deficit for $z = 11$ and between $10$ and $20\%$ for $z = 8$.

\subsection{Isocontour length}
\label{app:B:L}
\begin{figure*}
    \centering
    \includegraphics[width=0.73\textwidth]{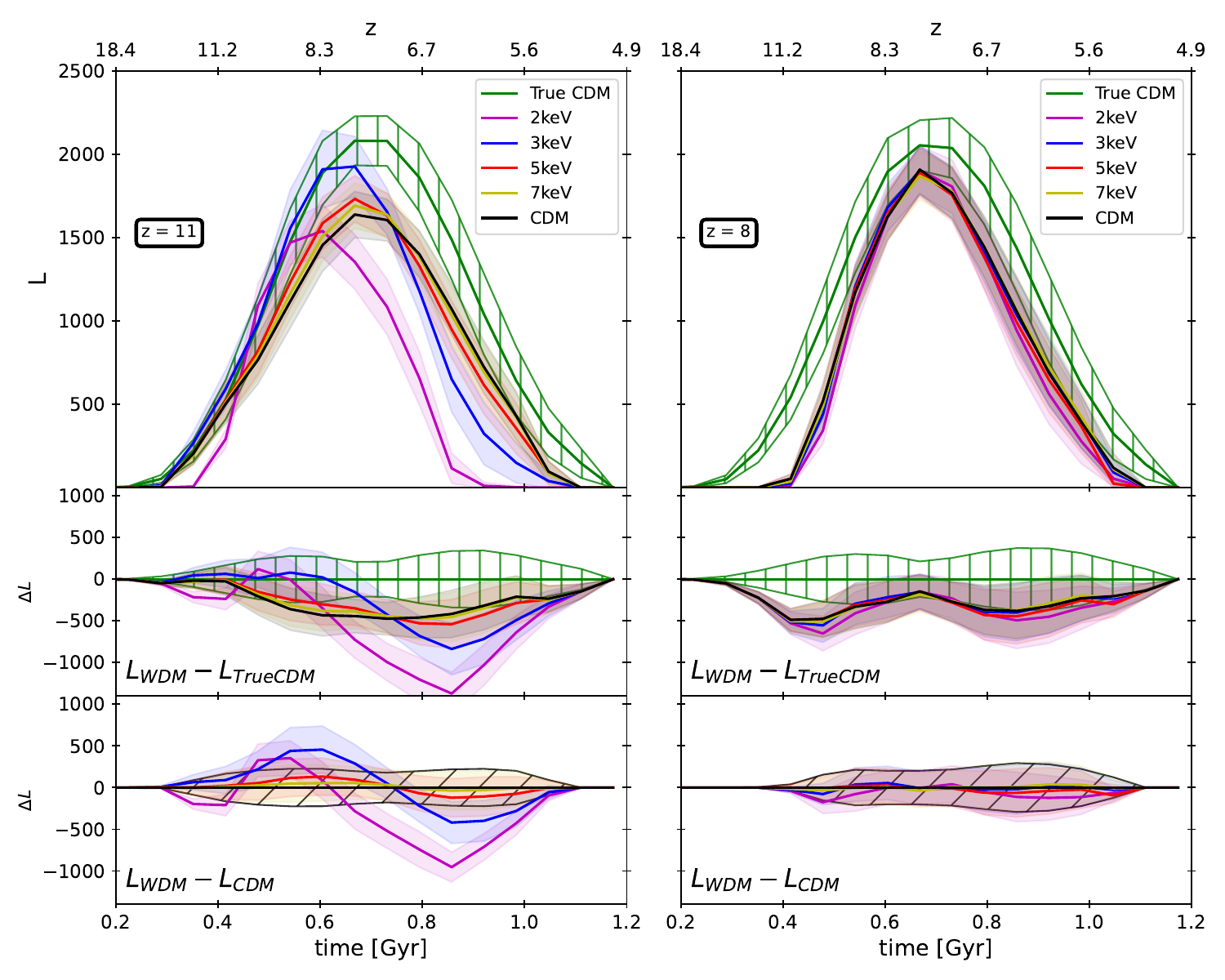}
    \caption{Total isocontour length $L$ for each predictor. Left and right are for $z=11$ and $8$ respectively. The top panel is the mean and standard deviation of $L$ with, in green, the ground truth coming from the CDM model. The middle panel represents the difference between WDM models and the ground truth, and the bottom panel represents the difference between WDM models and the CDM prediction.}
    \label{fig:L_vs_CDM}
\end{figure*}
Figure \ref{fig:L_vs_CDM} illustrates the total isocontour length, denoted as L, for each predictor, with the results for $z=8$ depicted on the right. At this redshift, every predictor reconstructs less length at each time compared to the ground truth. Additionally, the predictions are fairly consistent with each other, showing differences ranging from 0.5$\%$ to 6$\%$ within the time range t$\in$[0.54, 0.85] Gyr when compared to the CDM predictor. However, they all exhibit deficits ranging from $8\%$ to $9\%$ compared to the ground truth at 0.66 Gyr, where the peak value is observed. Notably, the initial sources ($< 0.5$ Gyr) are poorly represented, as predictors struggle to detect the seeds of reionisation. This challenge arises due to the presence of large ionised bubbles in the 21-cm maps at z=8, hiding the main seeds within their centres.
For $z=11$, the differences between predictors become more pronounced, with all predictors inferring an isocontour length significantly shorter than at z=8. Specifically, there is a 35$\%$, 8$\%$, 17$\%$, 19$\%$, 22$\%$ deficit for predictors 2, 3, 5, 7 keV, and CDM, respectively, at t=0.66 Gyr. However, in this scenario, predictors demonstrate better consistency with the ground-truth at earlier times ($< 0.5$ Gyr), falling within $1\sigma$. Surprisingly, the 3 keV predictor nearly perfectly reconstructs the isocontour length until 0.66 Gyr but deviates drastically afterwards, completely missing the isocontours at later times. 

\subsection{Minima - reionisation sources}
\label{app:B:N}
\begin{figure*}
    \centering
    \includegraphics[width=0.73\textwidth]{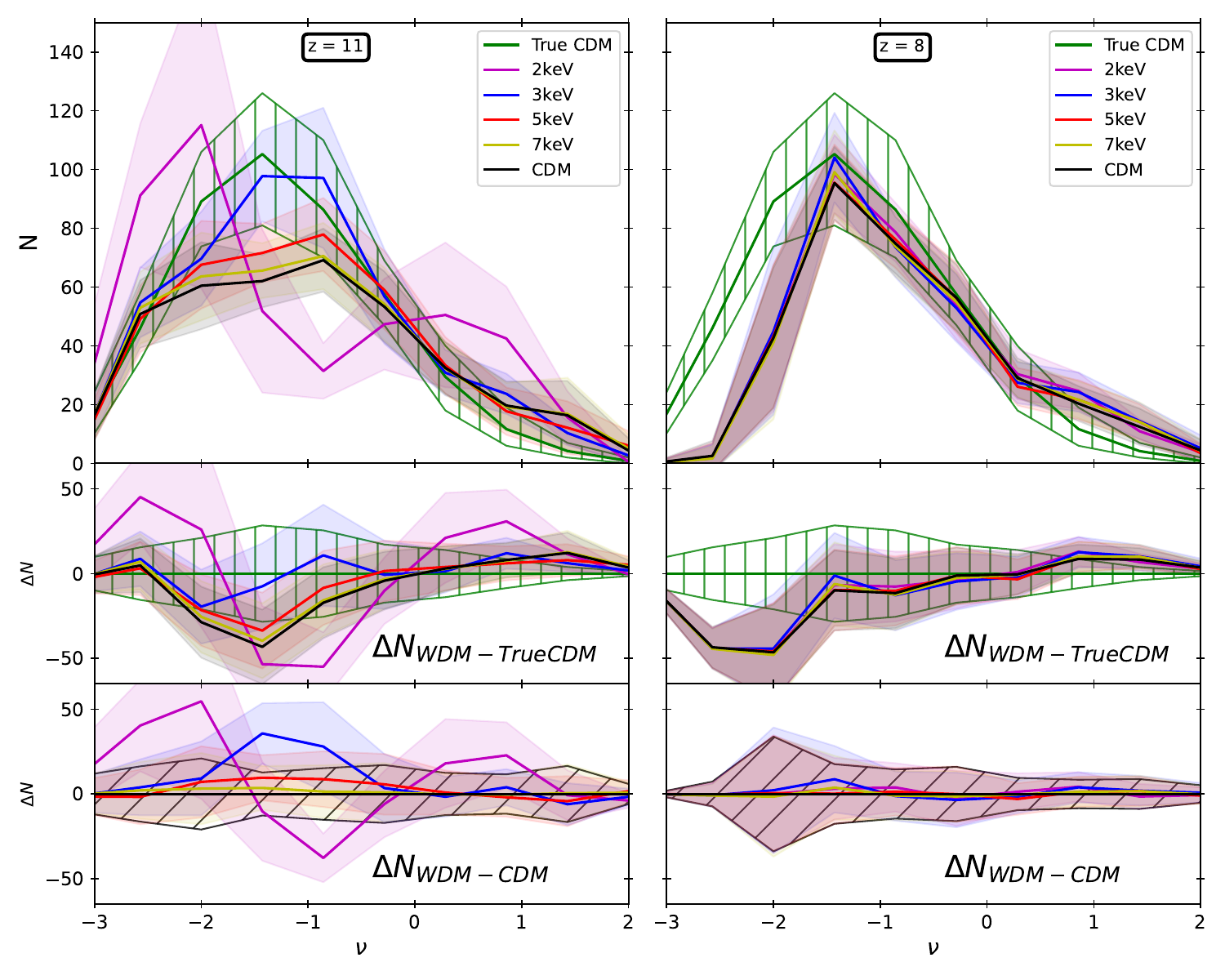}
    \caption{Minima number $N$ for each predictor. Left and right are for z=11 and 8 respectively. The top panel is the mean and standard deviation of $N$ with, in green, the ground truth coming from the CDM model. The middle panel represents the difference between WDM models and the ground truth, and the bottom panel represents the difference between WDM models and the CDM prediction.}
    \label{fig:N_vs_CDM}
\end{figure*}
Figure \ref{fig:N_vs_CDM} illustrates N for the $\TR$ predictions. The middle and bottom panels depict the differences between the True field (in green) and the CDM prediction (in black), respectively. For $z=11$, the 5, 7 keV and CDM predictors infer a $\TR$ field that closely aligns in terms of behaviour: they predict a consistent number of minima, indicating a similar number of reionisation seeds within the 21-cm signal at this redshift. On the other hand, these $z=11$ predictors tend to generate more minima at late time than the ground truth, seemingly hallucinating reionisation sources. At intermediate stages, where the distribution of minima reaches its maximum value ($\nu_{\TR}\approx -1.2$), these three predictors retrieve fewer minima than the ground truth: the ionised bubbles are few and describe the seeds of reionisation, yet the ionised regions corresponding to the intermediate sources are not evident in the 21-cm maps, concealing the source of UV radiation. It is noteworthy that our algorithm prioritises minimising the loss function and may favour recovering first sources to achieve the maximal global score (MSE in this case). 
The 2 keV predictor finds more minima at early times, as seen in Figure \ref{fig:Predictions_maps}: It constructs toroidal structures around reionisation sources, thereby boosting the number of minima found with this method. To understand why, a comparison between the 21-cm maps for CDM and the 2 keV maps is essential (see Figure \ref{fig:pred_right_obs}). In the 2 keV maps, most of the 21-cm map is seen with negative values (blue). However, the main interesting feature arises when moving out of ionised bubbles (white): there is a sharp turn to negative values. In contrast, in the CDM maps, there are positive values (in red) due to the heating of the IGM by galaxies. This phenomenon biases the 2 keV predictor, which struggles to analyse the mock CDM observation. On the other hand, at $z=8$, all the predictors broadly agree. Although they describe similar minima distribution, they fail to recover the number of early sources, where they exhibit a deficit compared to the ground truth. This result was anticipated since, at this redshift, the 21-cm maps have large HII regions hiding the sources somewhere in their centre, making the localisation of reionisation seeds challenging.

\section{WDM Baseline - 2 KeV}
\label{app:C:WDMref}
\begin{figure}
    \centering
    \includegraphics[width=0.5\textwidth]{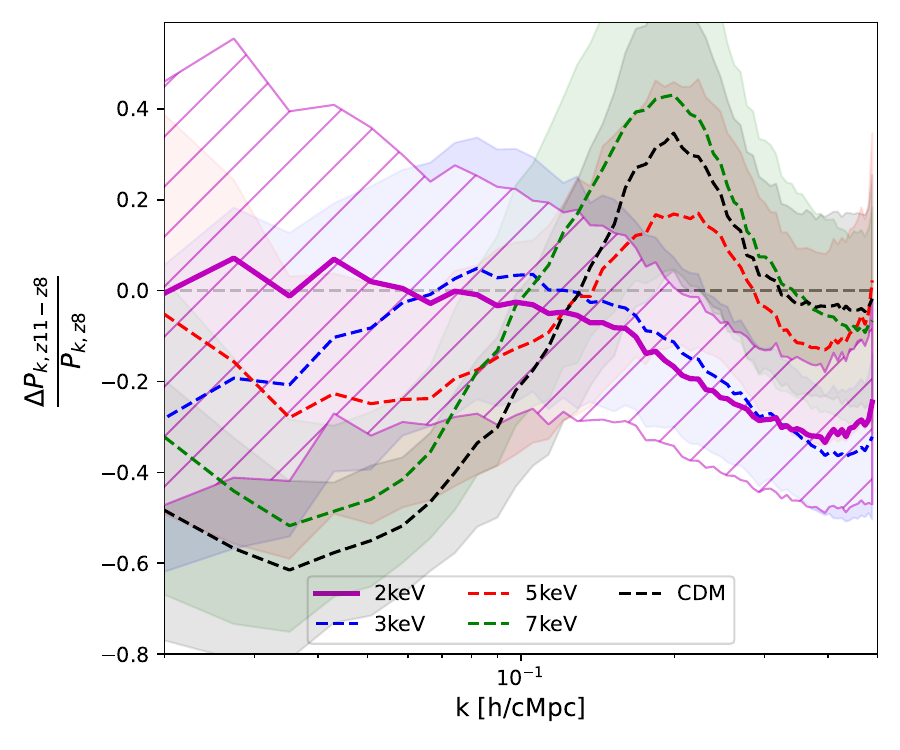}
    \caption{\textcolor{black}{Comparison of the $\TR$ power spectrum between redshift 11 and 8 for each model. Dashed/shaded areas stand for the standard deviation, while solid lines depict the mean. Here, the 2keV model represents the baseline.}}
    \label{fig:WDMref_pk}
\end{figure}
\textcolor{black}{In this section, we present a brief result where the 2 keV WDM model is selected as the mock observation instead of CDM. Figure \ref{fig:WDMref_pk} shows $\Delta P_k$, the relative difference in $P_k$ between $z=11$ and $z=8$, normalised by dividing by $P_k$ at $z=8$ (see Section \ref{subsec:Pk}).
In this case, the CNN trained on the 2 keV model (magenta curve) is taken as the baseline, and we observe a similar scenario to that found when CDM is used as the reference. The corresponding $R^2_\Delta$ values are 1.0, 0.53, -2.6, -7.5, and -7.0 for the 2 keV, 3 keV, 5 keV, 7 keV, and CDM models, respectively.
Here, only the 3 keV model appears to be comparable to the 2 keV case and could be considered valid. In contrast, due to their negative $R^2_\Delta$ values, the other three models are less likely acceptable and may be excluded.}

\section{Estimation of the network epistemic uncertainty }
\label{app:D:sigma_model}
\textcolor{black}{The epistemic uncertainty quantifies the uncertainty associated with the neural network parameters and reflects how confident the model is in its predictions. In this work, it is estimated using Monte Carlo (MC) dropout.
For each predictor model (2 keV, 3 keV, 5 keV, 7 keV, and CDM), we consider 50 independent 21-cm input maps (for $z=11$ and $z=8$) and perform 50 stochastic forward passes per map in inference mode with dropout enabled. In this configuration, the network remains stochastic, and each forward pass yields a different realisation of the predicted $\TR$ field. This implies that a single input map produces an ensemble of 50 distinct predictions, which we use to estimate epistemic uncertainty.
For each predicted realisation, we compute the derived observable of interest (e.g. $Q_{\mathrm{HI}}$), resulting in a set of 50 predictions per map and per redshift ($z=11$ and $z=8$). We then compute the coherence statistic between redshifts, defined as $z_{11} - z_{8}$, as illustrated in e.g. Fig.~\ref{fig:QHI11-QHI8}. For each input map, we take the mean and standard deviation of this statistic over the 50 stochastic forward passes.
Finally, the epistemic uncertainty of a given model is obtained by averaging the standard deviation over the 50 input maps.}

\end{appendix}

\end{document}